\documentclass[review]{elsarticle}

\usepackage{lineno,hyperref}
\usepackage{color}
\usepackage{graphicx}
\usepackage{graphics}
\usepackage{amsthm}
\usepackage{subfigure}
\modulolinenumbers[5]
\usepackage{soul}
\usepackage{blindtext}
\journal{Journal of \LaTeX\ Templates}

\begin{document}

\begin{frontmatter}

\title{The Northern and Southern Mid-latitude Ionospheric Trough using Global IGS vTEC maps}

\tnotetext[mytitlenote]{Fully documented templates are available in the elsarticle package on \href{http://www.ctan.org/tex-archive/macros/latex/contrib/elsarticle}{CTAN}.}

\author{Mar\'\i{}a Paula Natali, Juan Manuel Casta\~{n}o and Amalia Meza}
\address{Laboratorio de Meteorolog\'{i}a espacial, Atm\'osfera terrestre, Geodesia, Geodin\'{a}mica, dise\~{n}o de Instrumental y Astrometr\'\i{}a (MAGGIA), Facultad de Ciencias Astron\'omicas y Geof\'\i{}sicas (FCAG), Universidad Nacional de La Plata (UNLP), Paseo del Bosque s/n, B1900FWA, La Plata, Argentina

Consejo Nacional de Investigaciones Cient\'\i{}ficas y T\'ecnicas (CONICET), Godoy Cruz 2290, C1425FQB, Buenos Aires, Argentina}



\author[mysecondaryaddress]{Mar\'\i{}a Paula Natali \corref{mycorrespondingauthor}}
\cortext[mycorrespondingauthor]{Correspondig author}
\ead{paula@fcaglp.unlp.edu.ar}

\address[mymainaddress]{Paseo del Bosque  s/n, Buenos Aires, Argentina}

\begin{abstract}
The aim of this work is the analysis of the mid-latitude ionospheric trough (MIT) using the Global Ionospheric Maps from IGS (GIMs) during the solar minimum, year 2008. This study was performed for different local times, 22, 00, 02 and 04 LT on the Northern and Southern hemisphere simultaneously. In the two hemispheres the MIT show asymmetric pattern. The high-latitude troughs are clearly distinguished in autumn and winter. Another feature is the longitudinal development towards the west of the geomagnetic pole covering a wider area in the Northern Hemisphere. Five empirical reference models were tested and compared with the MIT minimum position obtained from GIMs at different local times for both hemisphere. The results show a better agreement with the observations for the Northern Hemisphere specially  with the K\"{o}ehnlein \& Raitt model. Fluctuations of 9 days and 27 days of the MIT minimum position are found, which could be related with the solar wind oscillations, especially for 00 and 02 LT in both hemisphere, suggesting a link between them. 
\end{abstract}

\begin{keyword}
\texttt{mid-latitude ionospheric trough, GIMs,  invariant latitude of MIT minimum position}\sep \LaTeX\sep Elsevier \sep template
\MSC[2010] 00-01\sep  99-00
\end{keyword}
\end{frontmatter}

\section{Introduction}
The electron density during nighttime is expected to decrease since its source of ionization, the solar extreme ultraviolet radiation, is absent. This nighttime behavior of the ionosphere is not so simple, there are a large number of anomalous phenomena that occur during the night under different conditions, i.e. the Weddell anomaly (Lin et al. 2010) (Meza et al. 2015), east-west difference (Natali \& Meza 2017) (Liu \& Yamamoto 2011), or the mid-latitude trough (Ishida et al. 2014) (Le et al. 2016). In particular, the mid-latitude ionospheric trough (MIT) is a depleted region of ionospheric plasma density in the F layer and lies just equatorward of the auroral equatorward boundary  (Yang et al. 2015).  One of the mechanism associated to the formation of the trough is the plasma stagnation and the decay in ionization during night hours in a region where coronation and convection of the electric fields acts  (Knudsen 1974). The MIT was observed and studied using different observations technique, from ground based to satellite (Krankowski et al. 2008) (Lee et al. 2011) (He et al. 2011).  

The occurrence of the MIT depends on latitude, longitude as well as the geomagnetic activity. There are some parameters which characterize the MIT, the trough minimum position, the trough depth and trough equatorial (or polar) half-width (Karpachev 2003) (Yang et al. 2015).

Rodger et al. (1992), Voiculescu et al. (2006) found that the MIT globally is more pronounced during autumn and winter months and less evident in spring and summer and in summer it is clearly observed near local midnight. Yang et al. (2015) performed a statistical analysis of the MIT over the Northern Hemisphere using GPS data during 2000-2014. They found that the trough minimum position depends primarily on geomagnetic activity, magnetic local time (MLT), and the season. They also conclude that trough depth depends on F10.7 and lesser with the MLT. 
The invariant latitude of MIT minimum is one of the more important parameter of the MIT location, and there are many models that consider the local time and geomagnetic activity to describe its variation (Kohnlein \& Raitt 1977) (Collis \& H\"{a}ggstr\"{o}m 1988) (Rycroft \& Burnell 1970) (Horvath \& Essex 2003) (Werner \& Pr\"{o}lss 1997).

Horvath \& Essex (2003) studied the diurnal, seasonal, spatial and magnetic activity variations of the southern-hemisphere mid-latitude trough using GPS and TOPEX satellite techniques during the low-sunspot number from February 1995 to February 1996. They found that the nighttime trough was well developed during summer for different geomagnetic activity. They also detected the trough twice in the Australian longitude region, being one at mid latitude and the other at higher latitudes. 

Lee et al. (2011) is the first paper studying the three dimensional structures of the MIT using GPS radio occultation experiment; they showed that the troughs in the two hemisphere are asymmetric and in Northern Hemisphere is more evident and stronger than  in the Southern Hemisphere during equinoctial seasons.
He et al. (2011) using GPS radio occultation experiment found at midnight the longitudinally deepest MIT occurs to the west of the geomagnetic pole in both the Northern and Southern Hemispheres during the equinox seasons and local summer. They conclude that the MIT location could be explained in terms of the neutral winds and the geometry of the magnetic field. This configuration produces that the plasma in the Southern Hemisphere goes downward where the declination is negative and viceversa. A similar situation occurred in the Northern Hemisphere. This effect enhanced the depletion. They also highlight a relationship between the  MIT  minimum position and the solar wind with an oscillation of 9 days. 

Since the establishment of the IGS tracking infrastructure, GNSS became a well-established tool for ionospheric sounding as these systems offer an unprecedented combination of accuracy, temporal and spatial resolution, and availability. In this structure the Ionosphere Working Group of the International GNSS Service (Iono-WG) created in 1998 (Feltens \& Schaer 1998) have had  the goal of generating reliable global vTEC maps (GIMs). GIMs offer a good opportunity to construct a better global climatology of ionospheric, especially in the Southern Hemisphere. (Yizengaw et al. 2005) used simultaneous global observations of the MIT from GIMs and the plasmapause position from IMAGE EUV, to analyze the correlation between MIT and plasmapause position, and they found an excellent agreement.

The objective of this work is to analyze the manifestation of MIT and the invariant latitude of MIT minimum is used for that purpose. The study is performed at different nighttime hours, i.e. premidnight and postmidnight for the Northern and Southern Hemisphere using GIMs products during 2008. The location of the MIT is compared with different empirical models and the relationship between this parameter and high-speed solar wind and geomagnetic activity is analyzed. First, the data used is described. Then, the results for different hours in both hemisphere are introduced along with a discussions. Finally, the conclusions are presented.

\begin{figure*}[htbp]
\centering
\subfigure[FMA]{\includegraphics[height=57mm, width=55mm]{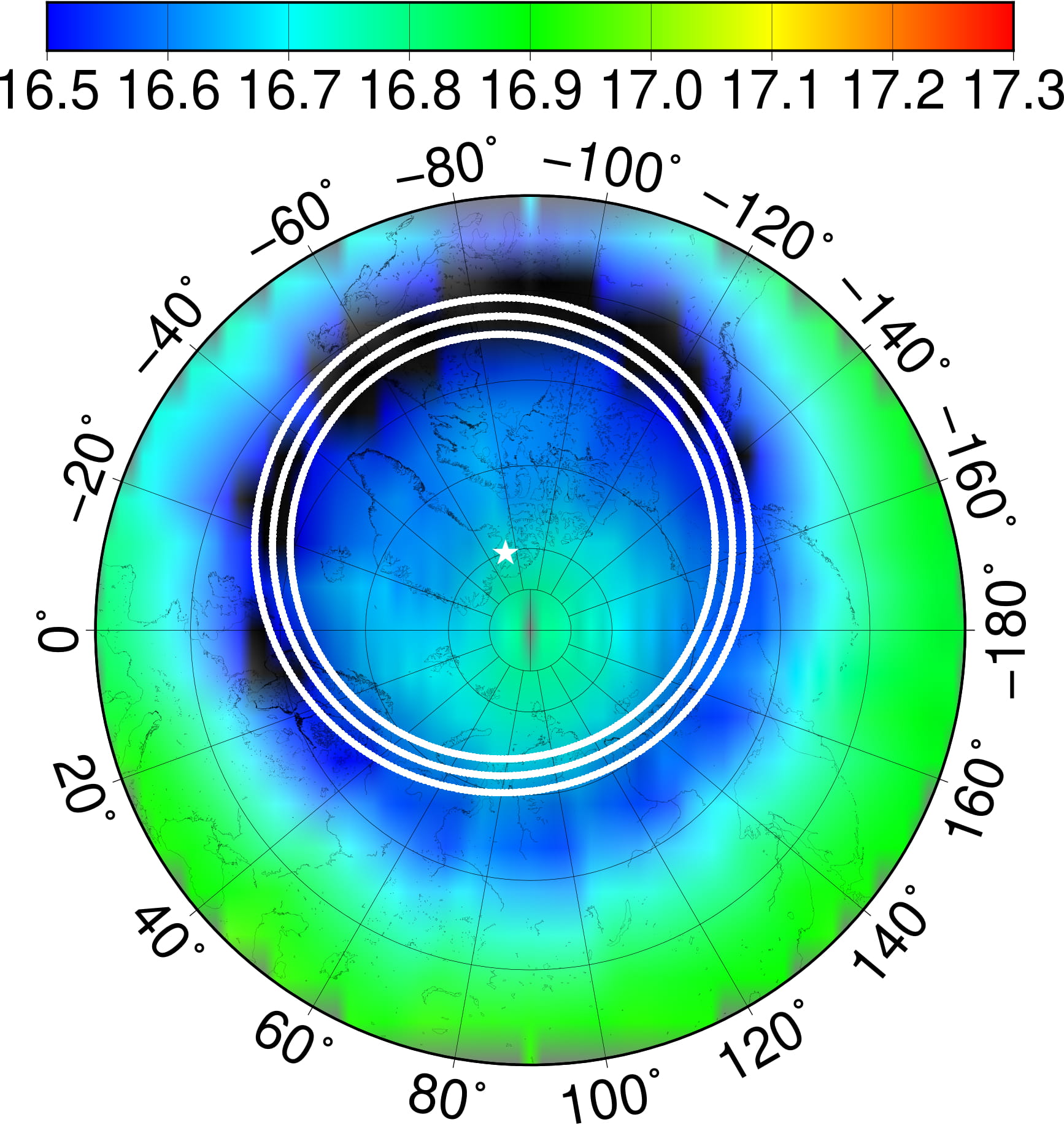}}\hspace{10mm}
\subfigure[MJJ]{\includegraphics[height=57mm, width=55mm]{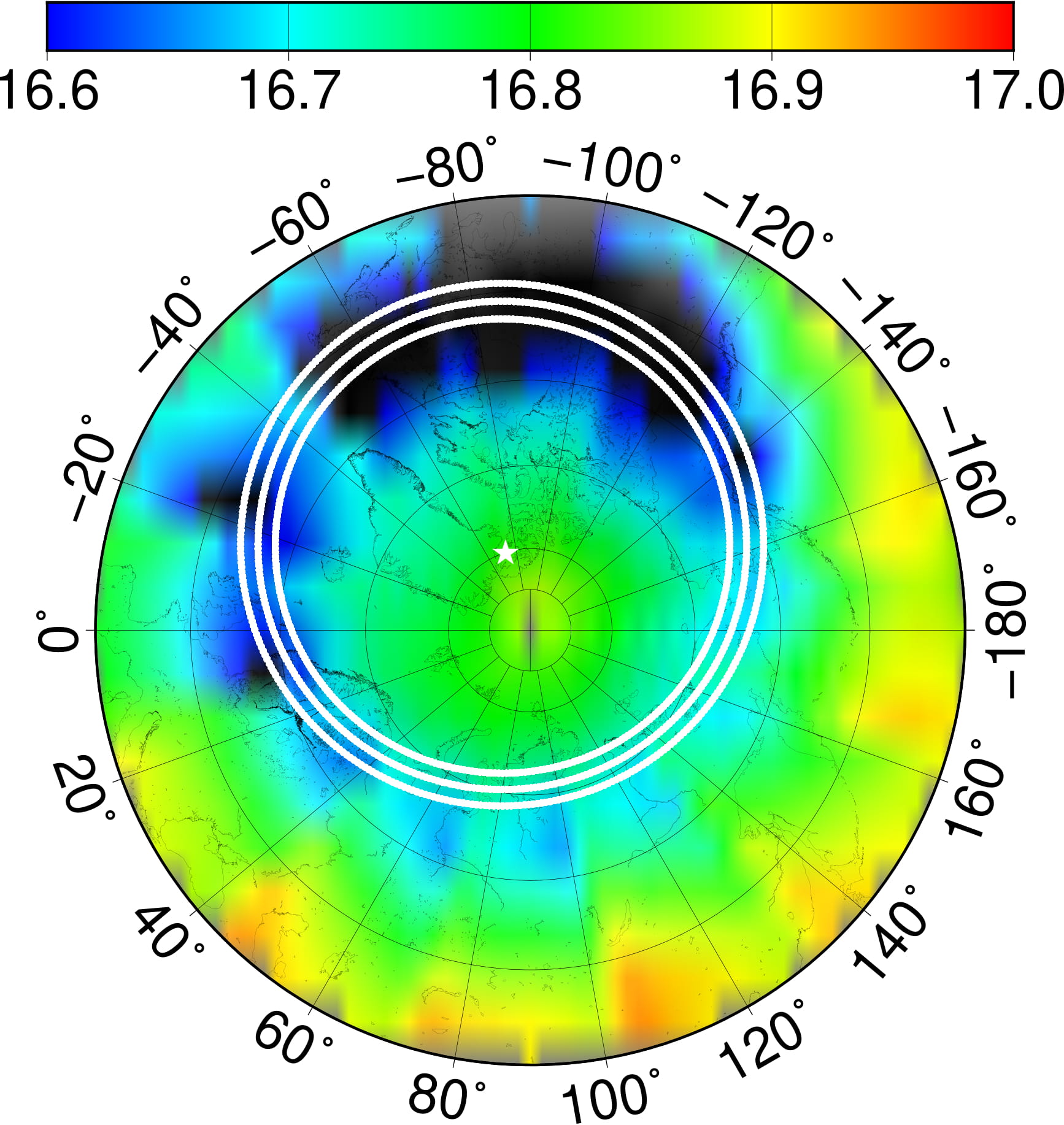}}\vspace{.5mm}
\subfigure[ASO]{\includegraphics[height=57mm, width=55mm]{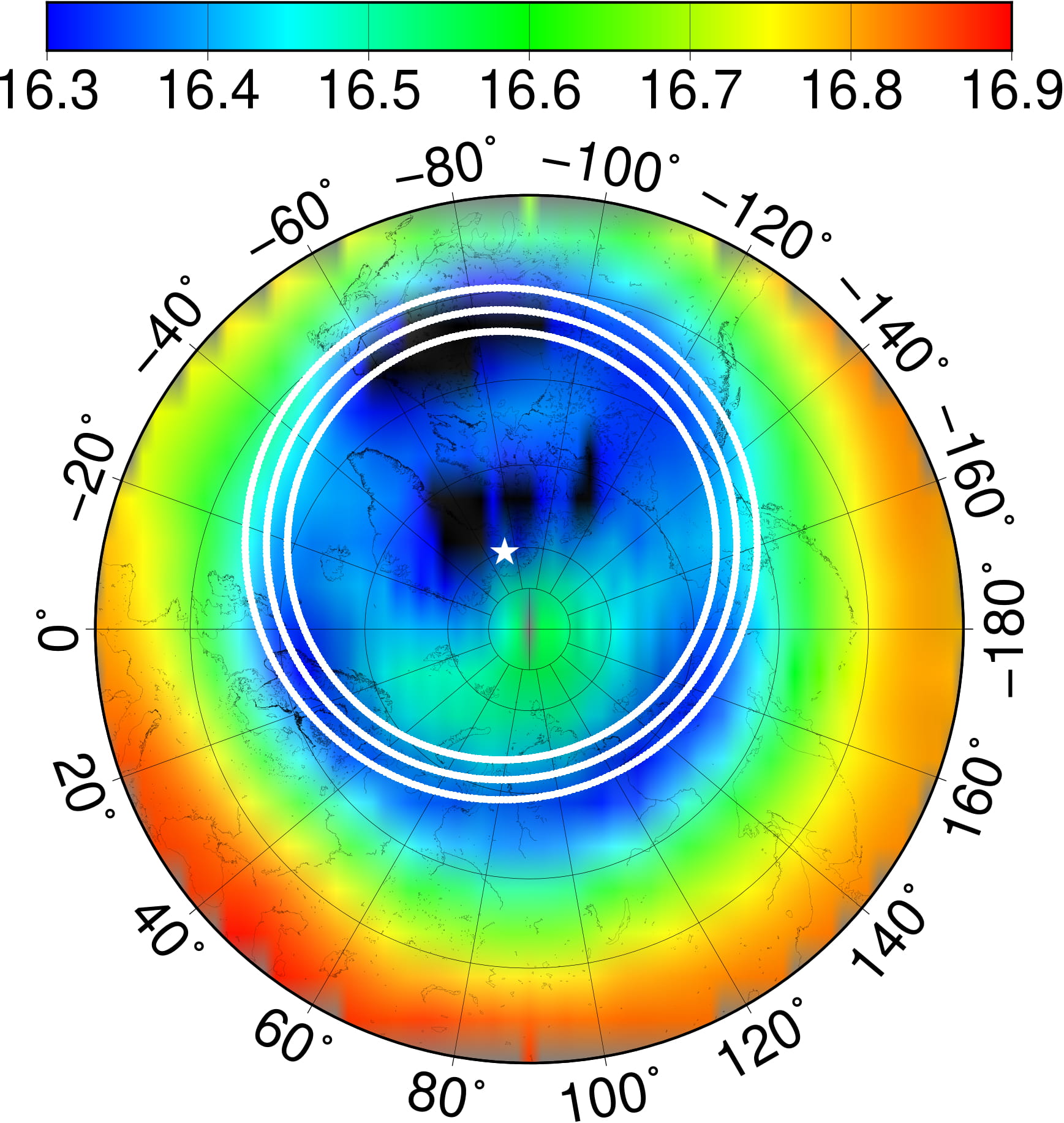}}\hspace{10mm}
\subfigure[NDJ]{\includegraphics[height=57mm, width=55mm]{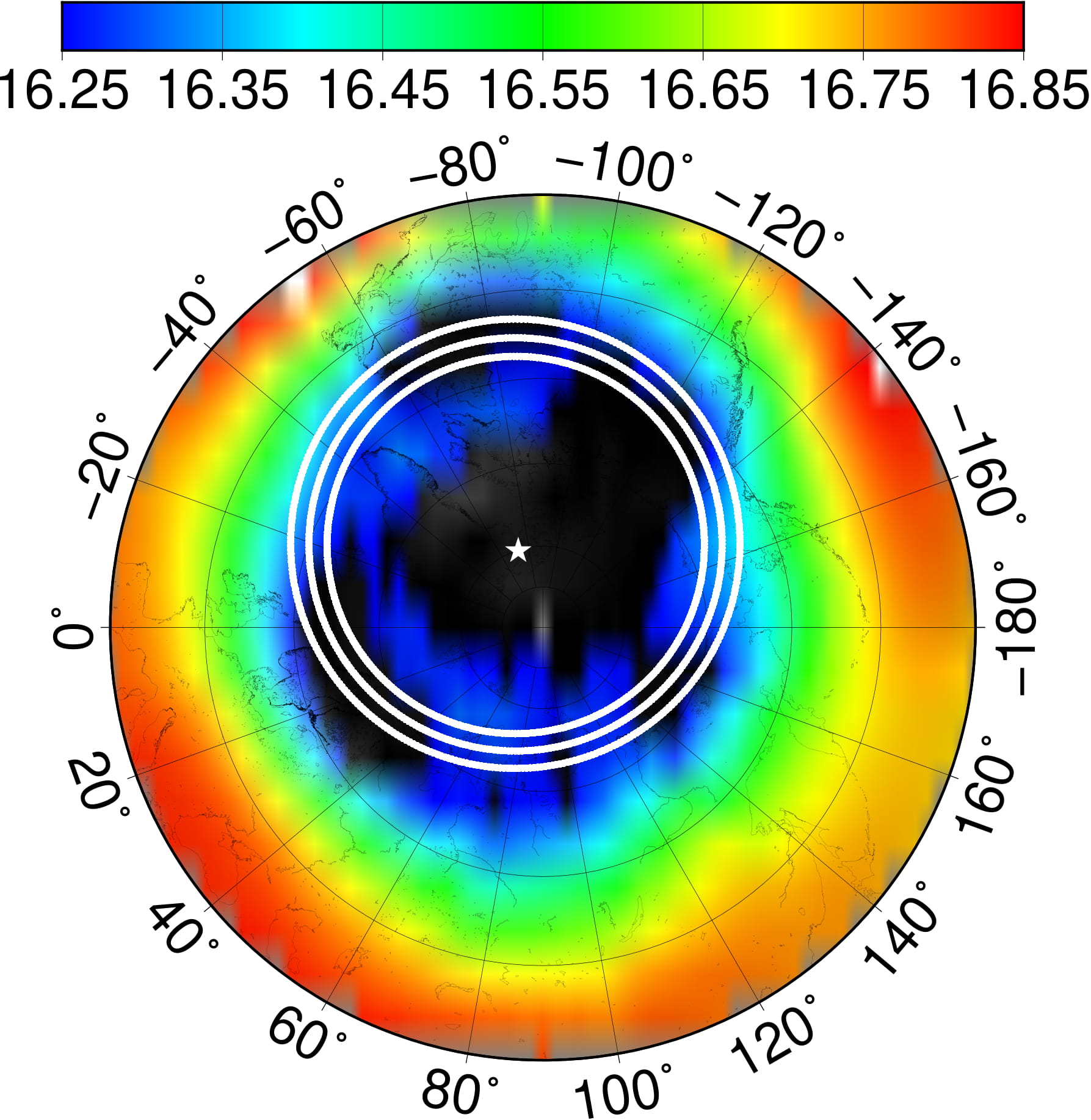}}\vspace{.5mm}
\caption{Seasonal mean vTEC maps at 00 LT for the four seasons for the Northern Hemisphere. In each map, the white solid lines show the geomagnetic latitude range where the  MIT minimum position is developed. In each map, the white star indicates the geomagnetic pole. Scale units are $\log _{10}(vTEC(TECU))$.} \label{Mapas_polaresHN}
\end{figure*}

\begin{figure*}[htbp]
\centering
\subfigure[FMA]{\includegraphics[height=57mm, width=55mm]{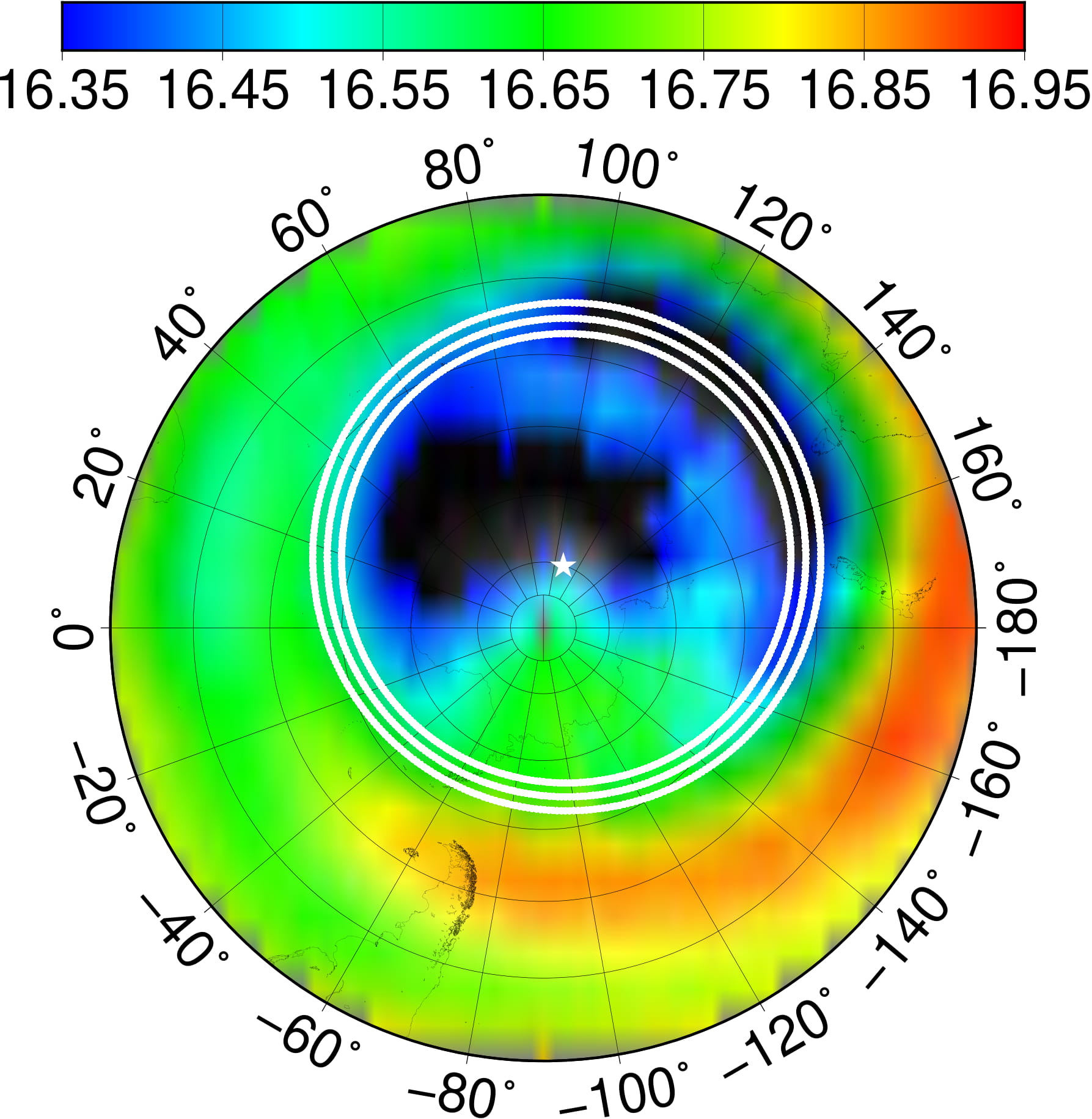}}\hspace{10mm}
\subfigure[MJJ]{\includegraphics[height=57mm, width=55mm]{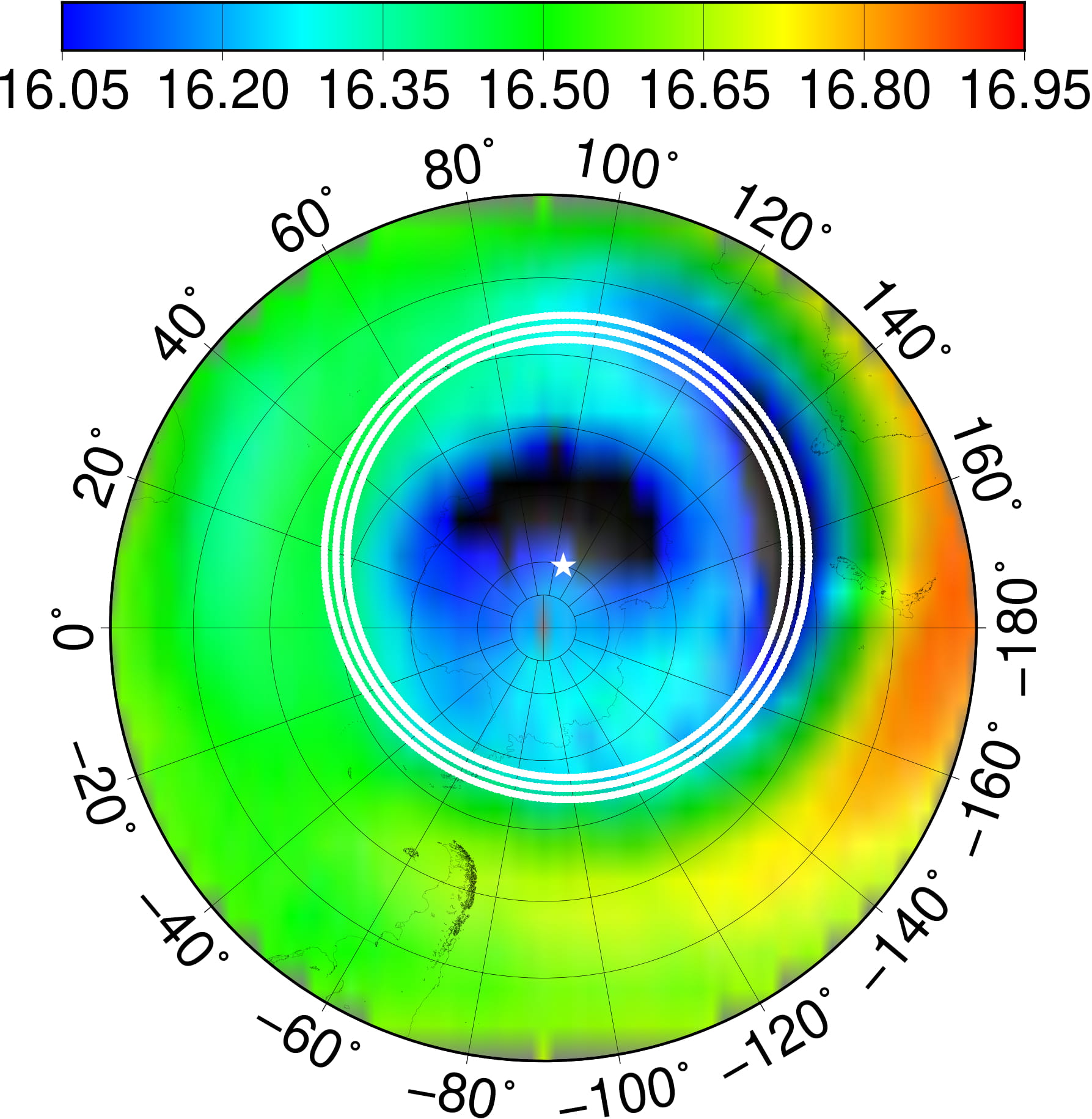}}\vspace{.5mm}
\subfigure[ASO]{\includegraphics[height=57mm, width=55mm]{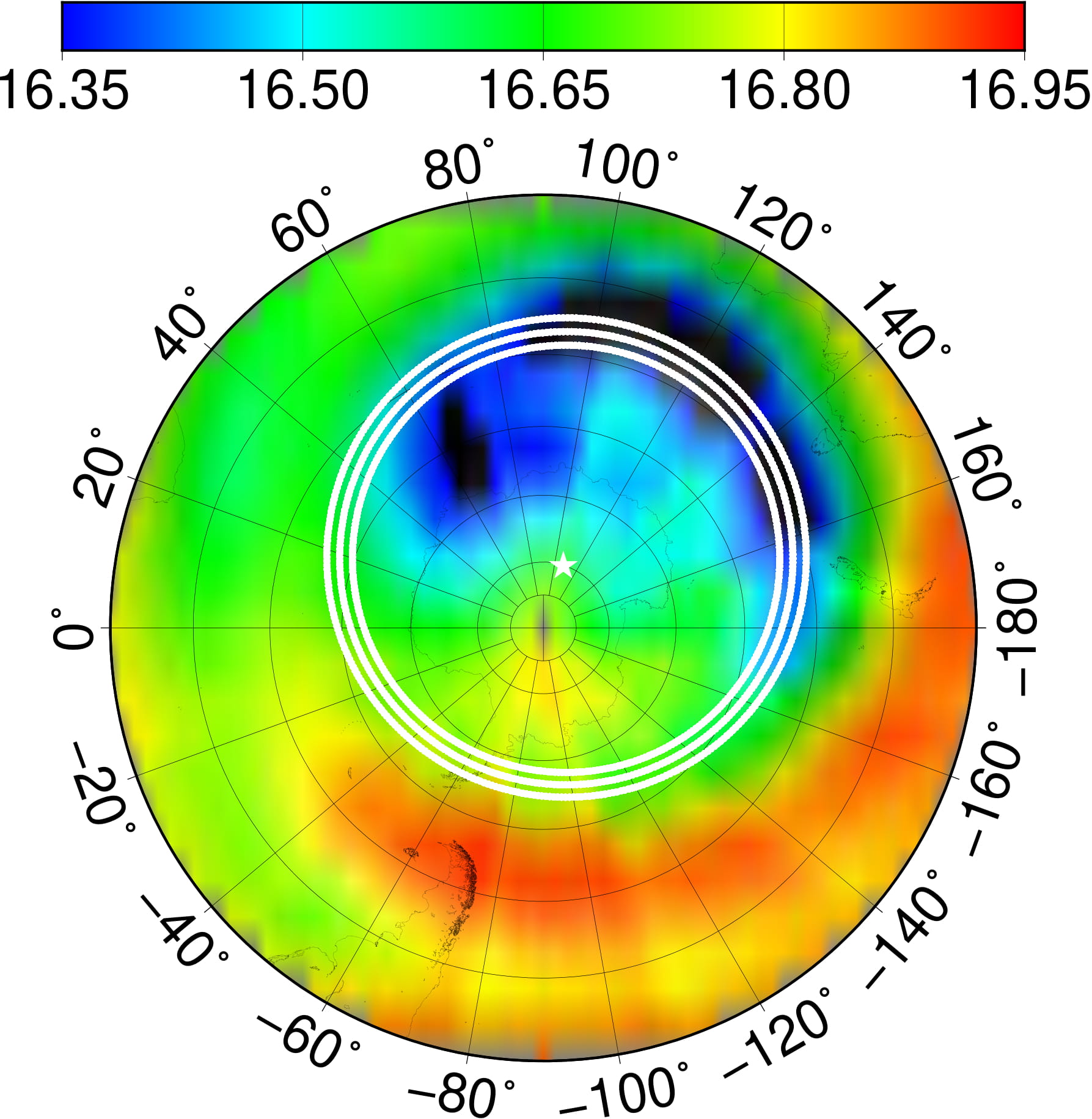}}\hspace{10mm}
\subfigure[NDJ]{\includegraphics[height=57mm, width=55mm]{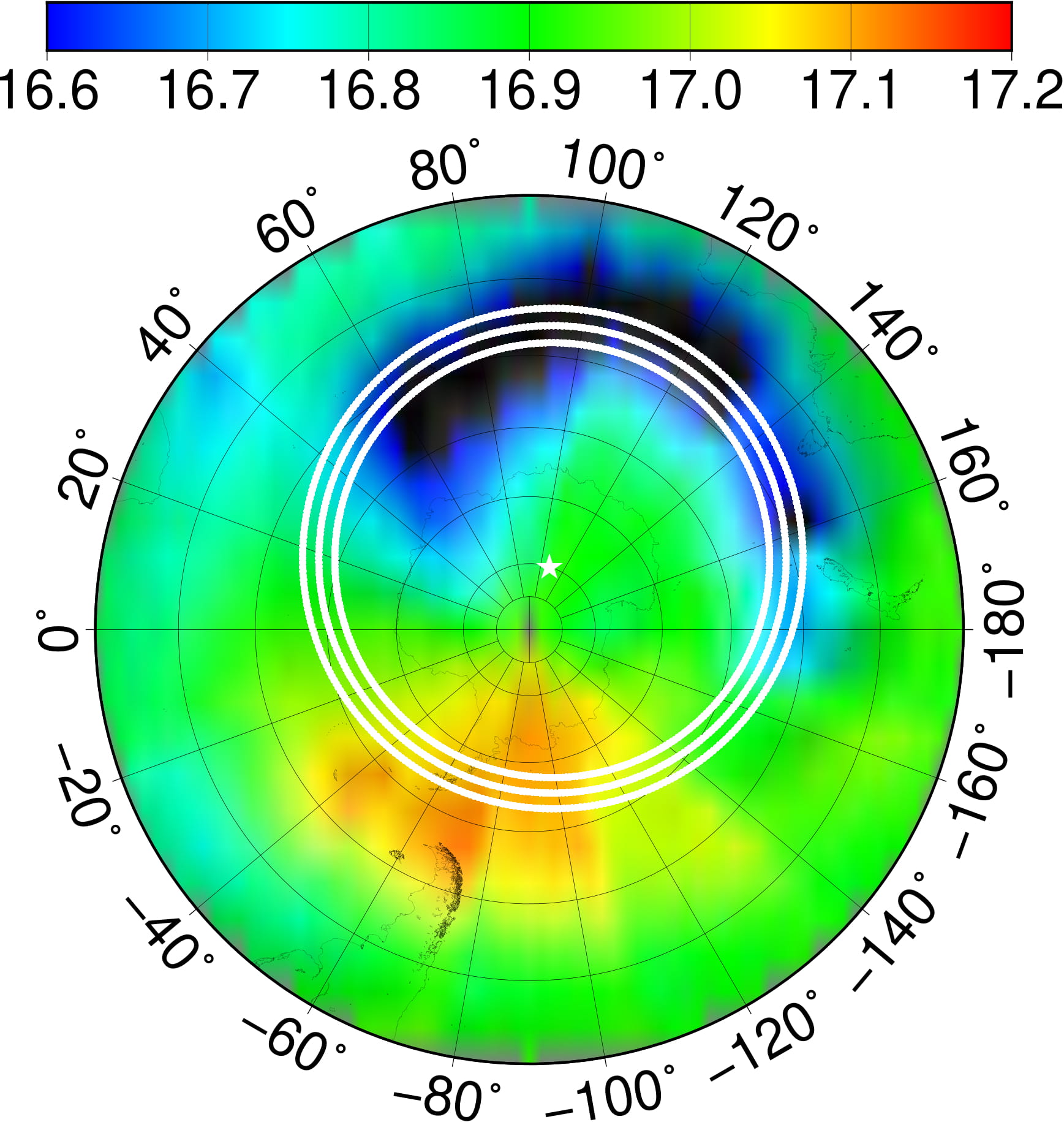}}\vspace{.5mm}
\caption{Seasonal mean vTEC maps at 00 LT for the four seasons for the Southern Hemisphere. In each map, the white solid lines show the geomagnetic latitude range where the  MIT minimum position is developed. In each map, the white star indicates the geomagnetic pole. Scale units are $\log _{10}(vTEC(TECU))$.} \label{Mapas_polaresHS}
\end{figure*}
\section{DATA}

Global vTEC maps from IGS were used in this work. These maps are a combined product which involve several steps: raw GNSS data measurements provided by the IGS GNSS ground network; an independent computation of vTEC maps by four analysis centers; evaluation and combination of the vTEC maps provided by the analysis centers. Finally the validation which is performed with an independent source of vTEC. More details about the generation of these maps could be found in (Hern\'andez Pajares et al., 2009). These maps provide ionospheric electron content information every 2 hours in a grid of 2.5 degrees in latitude and 5 degrees in longitude. GIM maps have a RMS error lower than 1 and 1.2 TECU $(1 TECU=10^{16}e^{-}m^{-2})$ for the Northern and Southern Hemisphere respectively. 
These maps were also compared with VTEC values from Topex and Jason altimetric satellites measurements and the RMS values obtained were 3.2 and 4.5 TECU in the Northern and Southern Hemisphere respectively for invariant geomagnetic latitudes higher than $60^{\circ}$ (Hern\'andez Pajares, 2004). 
In order to represent different local time, the GIMs were ordered to obtain 1 map per day at 22 LT, 00 LT, 02 LT, 04 LT for the year 2008. The reordering scheme was as follows: assuming that the ionosphere does not change in a two-hour window, we considered slices of 30 degrees each centered on the same local time, from all the maps for each day. Then, the slices were merged into a new vTEC map corresponding to their central longitude. Thus, following this procedure a new vTEC map was built matching to the local time selected.

The relationship between the geomagnetic latitude of MIT minimum and solar behavior is analyzed using the high-speed solar wind  ($v_{ws}$) which is obtained from the Advanced Composition Explorer (ACE, http://www.srl.caltech.edu /ACE/ASC/level2). This is an Explorer mission that was managed by the Office of Space Science Mission and Payload Development Division of the National Aeronautics and Space Administration (NASA). The data was hourly average. Outliers were replaced using a linear interpolation. 

Finally, the \(Kp\) geomagnetic index is used to study the relationship, between the MIT minimum position and geomagnetic activity. This index is obtained every hour from NASA product (https://omniweb.gsfc.nasa.gov/form/dx1.html).

\section{RESULTS AND DISCUSSIONS}

The results were organized in:  FMA (February, March and April), MJJ (May, June and July), ASO (August, September and October) and NDJ (November, December and January). Figures \ref{Mapas_polaresHN} and \ref{Mapas_polaresHS} show the maps of the seasonal mean values of vTEC at midnight (for the other local time the results are similar but the figures are not included in this manuscript). The trough is plainly seen in both hemisphere, showing a different behavior for each season. There is a clear asymmetry between both hemisphere. For both Hemisphere the high-latitude trough is also evident during autumn and winter (Grebowsky et al. 1983). In all cases the MIT minimum position is well aligned with the geomagnetic latitude $+62^{\circ}$ and $-55^{\circ}$, white central line in Figs. \ref{Mapas_polaresHN} and \ref{Mapas_polaresHS}, for the Northern and Southern Hemisphere respectively. The MIT shows a strong longitudinal structure toward west of the northern and southern geomagnetic pole (white star). In the Northern Hemisphere this structure covered a wider longitudinal area than those in the Southern Hemisphere, one reason could be the nighttime plasma density enhancement of the Weddell Sea Anomaly. Similar results are shown by (He et al. 2011).

\begin{figure*}[ht]
\raggedright
\includegraphics[width=1.1\textwidth]{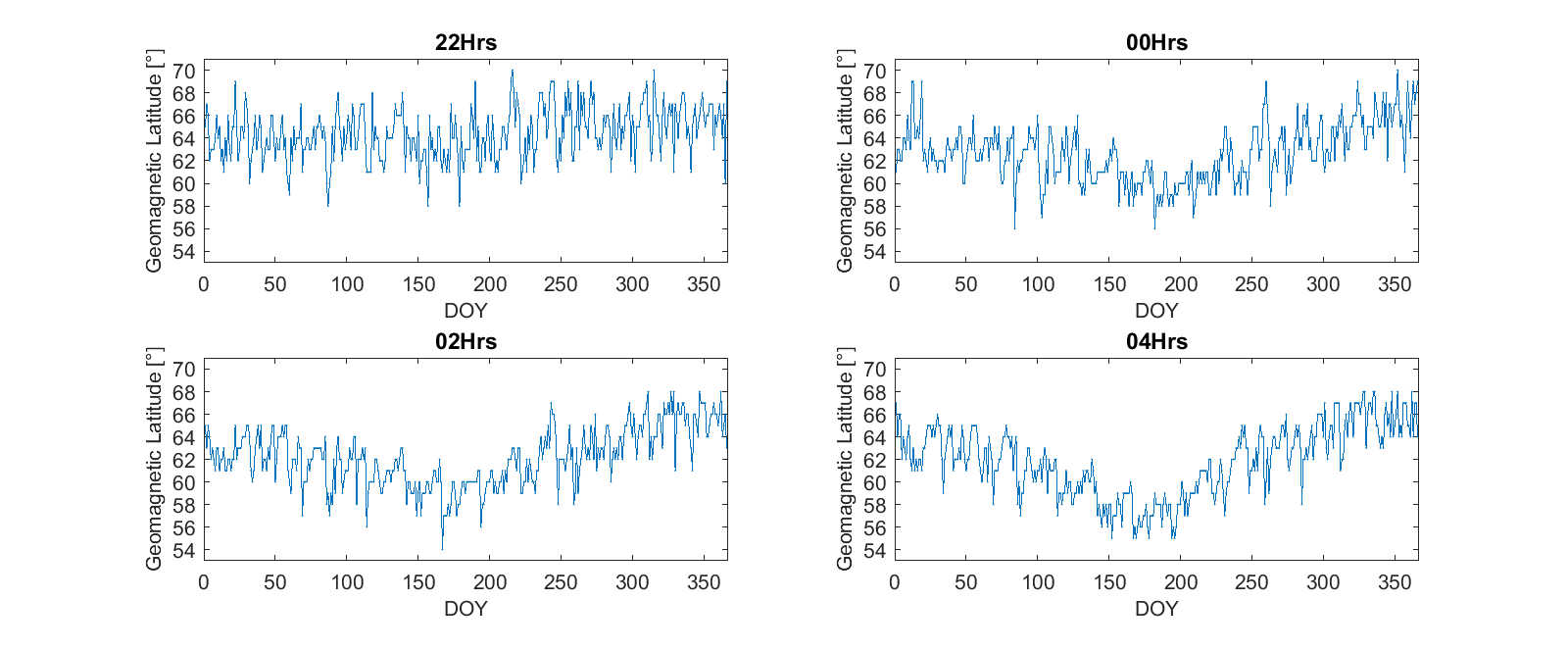}  
\caption{MIT minimum position for the Northern Hemisphere during 2008 at different local time.} \label{HN_MIT}
\end{figure*}

\begin{figure*}[ht]
\includegraphics[width=1.1\textwidth]{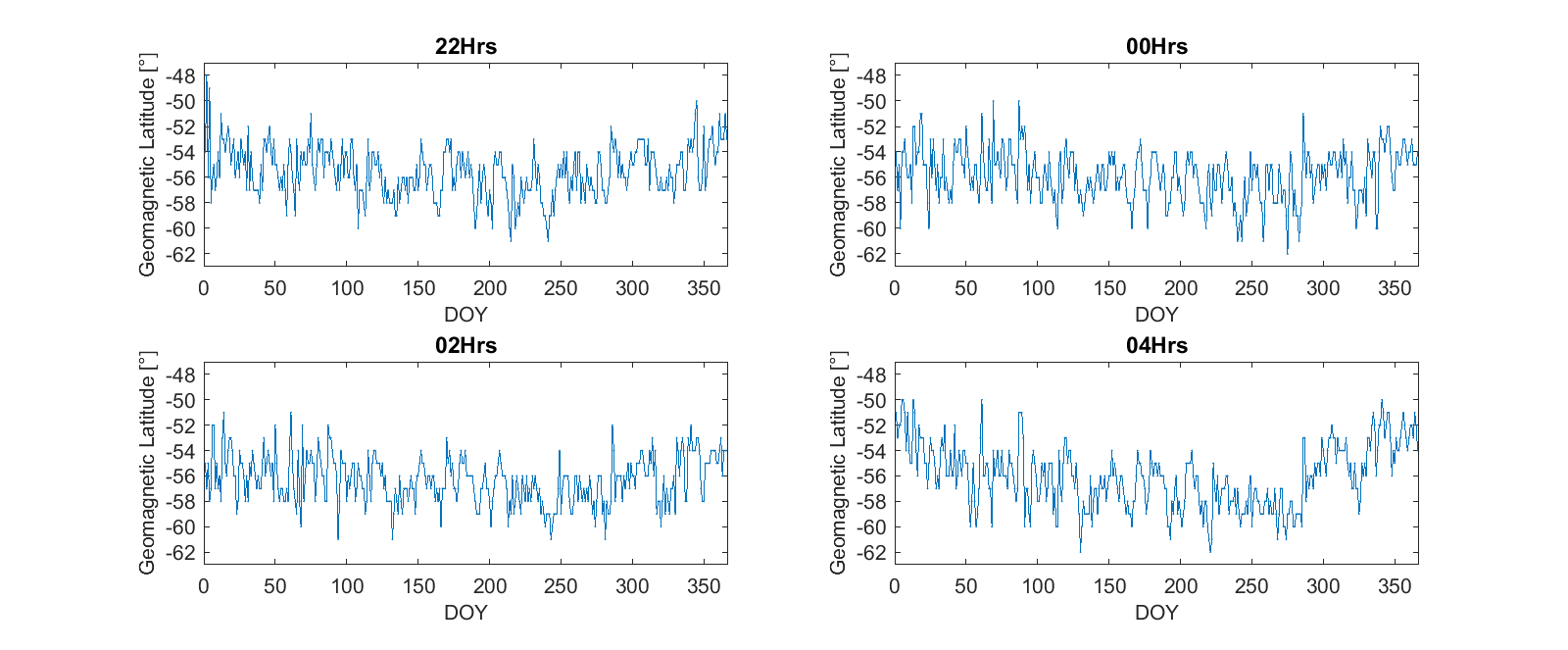} \centering
\caption{MIT minimum position for the Southern Hemisphere during 2008 at different local time.} \label{HS_MIT}
\end{figure*}

Figures \ref{HN_MIT} and \ref{HS_MIT} show the MIT minimum position for the year 2008, analyzed at different local times and both hemispheres. The procedure to obtain the MIT minimum position is as follows: longitudinal limits are selected considering the sectors where the mid-latitude ionospheric trough persists throughout the year. Consequently, longitudes between $-170^{\circ}$ and $-100^{\circ}$; and $90^{\circ}$ and $170^{\circ}$ for the Northern and Southern Hemisphere respectively were selected to obtain the geomagnetic latitudinal mean vTEC value for each day. Then, the daily geomagnetic latitude of the MIT minimum position is obtained by fitting to the mean vTEC values using a fourth degree polynomial. For the Southern Hemisphere the region of the Weddell Sea Anomaly was excluded (Lee et al. 2011).

From these figures it is possible to see how the position varies in latitude, along the year for the different local time. In both hemisphere the MIT minimum position is closest to the equator during local summer (Horvath and Essex, 2003).

Muldrew (1965) is the first paper to relate the MIT minimum position shift equatorward with an increase in geomagnetic activity (Kp index). Consequently, several empirical models were developed to determine the invariant geomagnetic latitude of the MIT minimum position and the level of geomagnetic activity. In this work five empirical models are used to determine the invariant latitude of the MIT minimum position as a function of the geomagnetic activity. Three of them use Kp index and the others AE index. A brief description of these models are presented along with their equations. 
 
Rycroft \& Burnell (1970), Collis \& H\"{a}ggstr\"{o}m (1988) and (K\"{o}ehnlein \& Raitt (1977) performed a statistical analysis of the main trough MIT location obtaining an empirical model. They obey a linear equation whose variables are \(Kp\) and local time (\( t\)). This equation implies a constant equatorward motion independent of both, activity level and local time (Eq.(\ref{Eq_1})). Each model use a simple multi-regression analysis of the data, being the linear equation  

\begin{equation}
\Lambda =A_0+a Kp + b t\
\label{Eq_1}
\end{equation}

where \(\Lambda\) is the trough MIT minimum position (invariant latitude in degree), \(A_0\)
the constant, \( a\) and \( b\) are the multi-regression coefficients and \(t\) is the time in hours from local midnight (positive after midnight and negative before midnight). Table \ref{TAB:stats} shows the constants and coefficients obtained by the authors mentioned before.

The remaining models are proposed by Werner and Pr\"{o}lss (1997). They develop two empirical models whose variables are \(AE\) and magnetic local time (\( MLT\)) to describe the position of the trough. Both models are, 
{
\begin{eqnarray}
& \Lambda   =  67.52 - 10.07 \cdot \cos \left ( \frac{2\pi}{24}\cdot MLT \right ) - 1.61 \cdot \sin \left ( \frac{2\pi}{24}\cdot MLT \right ) \nonumber \\
& - 1.78 \cdot \sin \left ( \frac{4\pi}{24}\cdot MLT \right )  +   3.06 + 1.61 \cdot \cos \left ( \frac{2\pi}{24}\cdot MLT \right ) + AE_{6} \cdot\\ 
&  \left [ -1.19 \cdot 10^{-2} - 4.45 \cdot 10^{-3} \cdot \cos \left ( \frac{2\pi}{24}\cdot MLT \right ) + 4.18 \cdot 10^{-3} \cdot \sin \left ( \frac{2\pi}{24}\cdot MLT \right) \right]  \nonumber
\end{eqnarray}

}
\begin{eqnarray}
 &   \Lambda = 59.18 - 1.129 \cdot t + 0.1197 \cdot t^{2} + 7.715 \cdot 10^{-3} \cdot t^{3}+ 2.88   \nonumber \\
 & + 1.23 \cdot \cos \left ( \frac{2\pi}{24}\cdot MLT \right ) - 0.50 \cdot \sin \left ( \frac{2\pi}{24}\cdot MLT \right ) + AE_{6} \cdot \\
 &\left [-1.25 \cdot 10^{-2} - 4.61 \cdot 10^{-3} \cdot \cos \left ( \frac{2\pi}{24}\cdot MLT \right ) + 2.80 \cdot 10^{-3} \cdot \sin \left ( \frac{2\pi}{24}\cdot MLT \right ) \right ] \nonumber
\end{eqnarray}

where \(\Lambda\) is the invariant latitude of  trough minimun position, MLT is the magnetic local time, $MLT=t+24$ before 0 MLT and $MLT=t$ after 0 MLT; and $AE_{6}$ is 

\begin{equation}
AE_{6}=\sum_{i=0}^{6}AE(x)\cdot {e^{-i/6}}/4.49
\end{equation}

with  $AE(x)$ the hourly averaged auroral electrojet index at time  $x= UT - i$. Eq. 2 and 3 are referred as model A and B respectively.  

The empirical relation Eq.(\ref{Eq_1}) derived by Rycroft \& Burnell (1970) used data collected during April-August 1963 (a period of low sunspot activity). Collis \& H\"{a}ggstr\"{o}m (1988) fitted the coefficient of Eq.(\ref{Eq_1}) using data collected during 1986-87 (a period of low solar activity). Finally, K\"{o}ehnlein \& Raitt (1977) used observations from the ESRO 4 satellite during November 1972 to April 1974, a period of relatively quiet solar conditions. The latter concluded that the formula is valid for equinox and winter. Werner \& Pr\"{o}lss (1997) models, Eqs. (2) and (3), are based on electron density measurements of the DE2 satellite which cover the time period August 1981 to February 1983.

Rodger \& Pinnock (1980) showed that the movement of the poleward edge of the trough does not obey this simple relationship Eq.(\ref{Eq_1}), and other authors like Collis \& H\"{a}ggstr\"{o}m (1988) confirm that neither does the trough minimum position. Werner \& Pr\"{o}lss (1997) found an agreement with Collis \& H\"{a}ggstr\"{o}m (1988) and K\"{o}ehnlein \& Raitt (1977) models; they also found a weak seasonal dependence of MIT minimum position. 

These models are compared with the respective MIT minimum position calculated in our work.  
In the figures the models are classified as follow: Rycroft (Rycroft \& Burnell 1970), K\"{o}ehnlein (K\"{o}ehnlein \& Raitt 1977), Collis (Collis \& H\"{a}ggstr\"{o}m 1988), W\&P$_{A}$ and W\&P$_{B}$ (Model A and Model B from Werner \& Pr\"{o}lss (1997) respectively). The error of the empirical invariant latitude estimation is between 2 and 3.5 degrees.
 
\begin{table}[ht]
\centering
\caption {Constant and coefficients values for the different models (invariant latitude).}
\label{models}
\begin{tabular}{|p{0.2\columnwidth}|p{0.15\columnwidth}|p{0.15\columnwidth}|p{0.15\columnwidth}|} \hline
\textbf{$Model $}  &  \textbf{$A_0$ } & \textbf{$a$} & \textbf{$b$ }  \\ \hline
Rycroft & 62.7 & -1.4  & -0.7  \\ \hline  
K\"{o}ehnlein & 65.2 & -2.1  & -0.5   \\ \hline  
Collis & 62.2 & -1.6  & -1.35   \\ \hline   
\end{tabular}
\label{TAB:stats}
\end{table}

\begin{figure*}[ht]
\includegraphics[height=110mm, width=120mm]{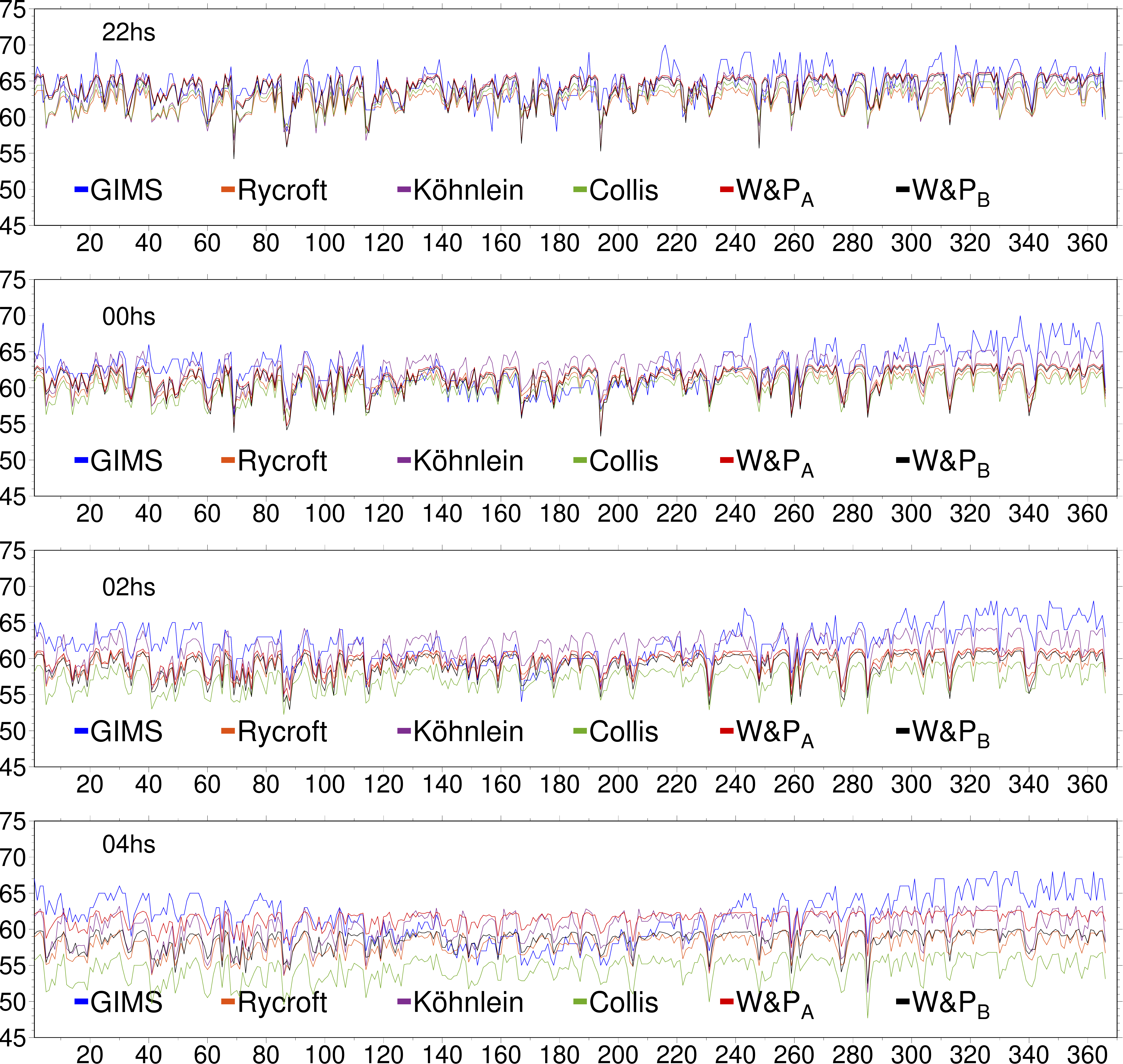} 
\caption{MIT minimum position from GIMs and invariant geomagnetic latitude from the six empirical models for the Northern Hemisphere.}
\label{GIM_ModelsHN}
\end{figure*}

\begin{figure*}[ht]
\includegraphics[height=110mm, width=120mm]{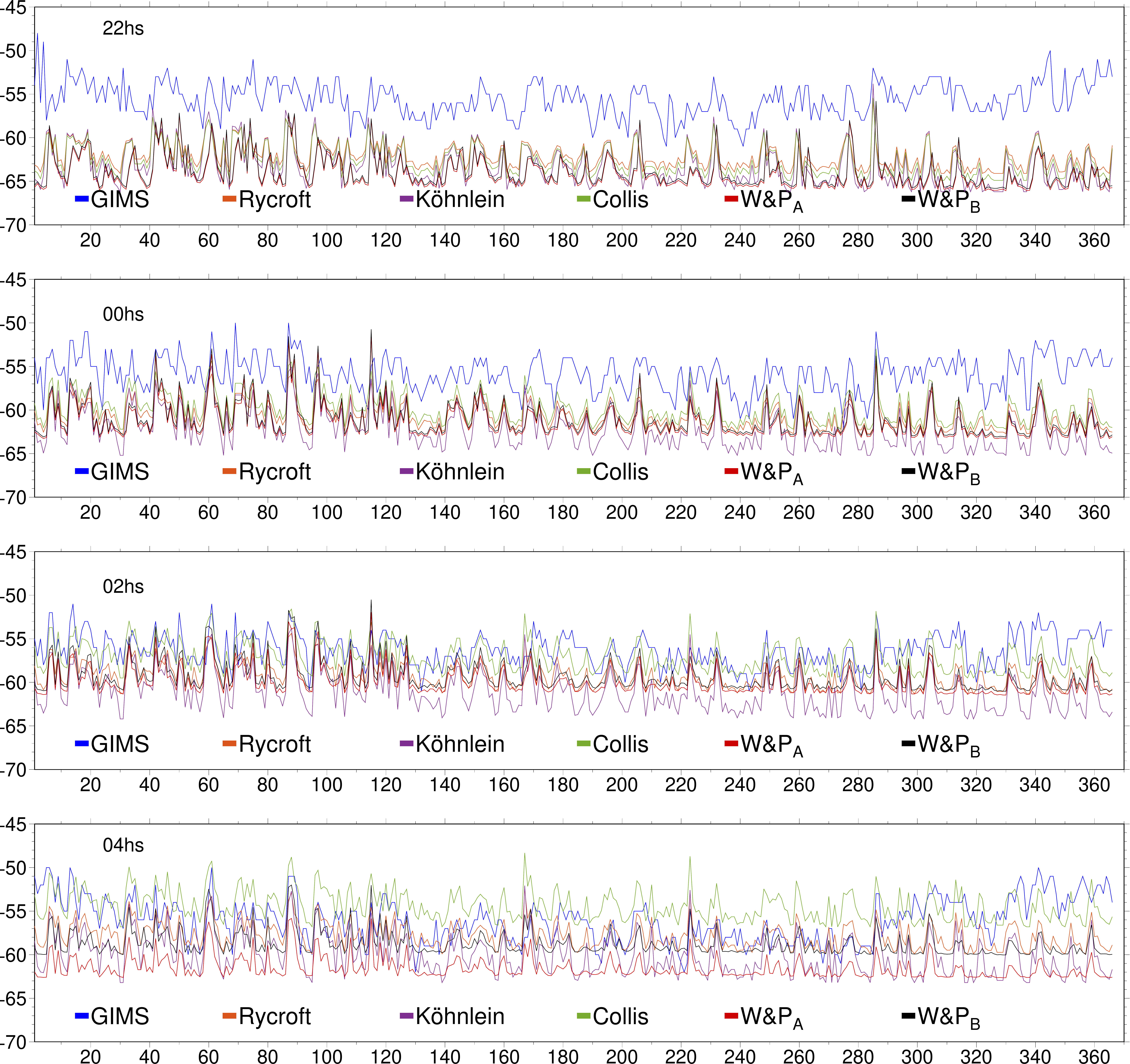} 
\caption{MIT minimum position from GIMs and invariant geomagnetic latitude from the six empirical models for the Southern Hemisphere.}
\label{GIM_ModelsHS}
\end{figure*}

\begin{figure*}[htbp]
\centering
\subfigure[]{\includegraphics[height=54mm, width=58mm]{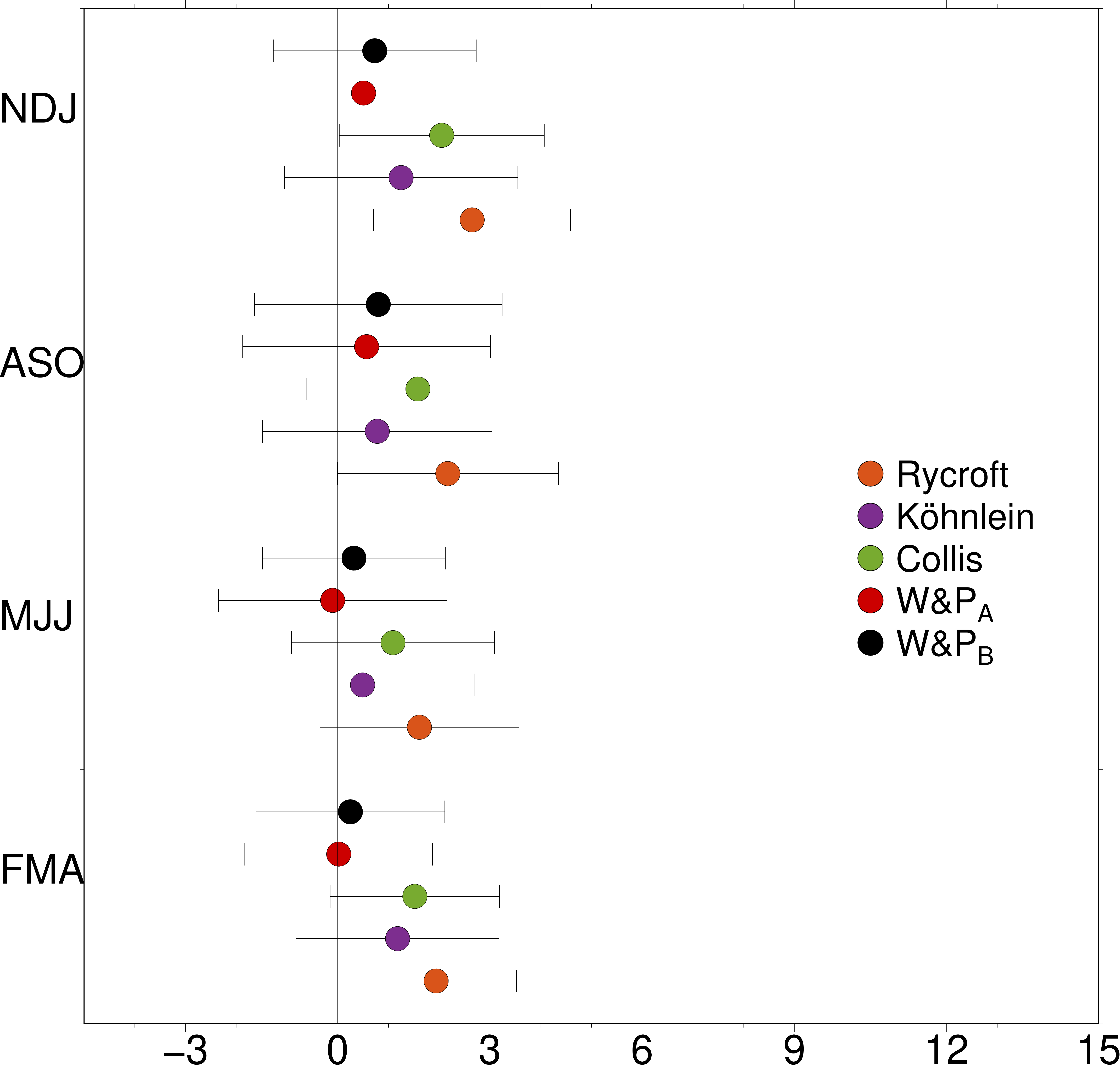}}\hspace{2mm}
\subfigure[]{\includegraphics[height=54mm, width=58mm]{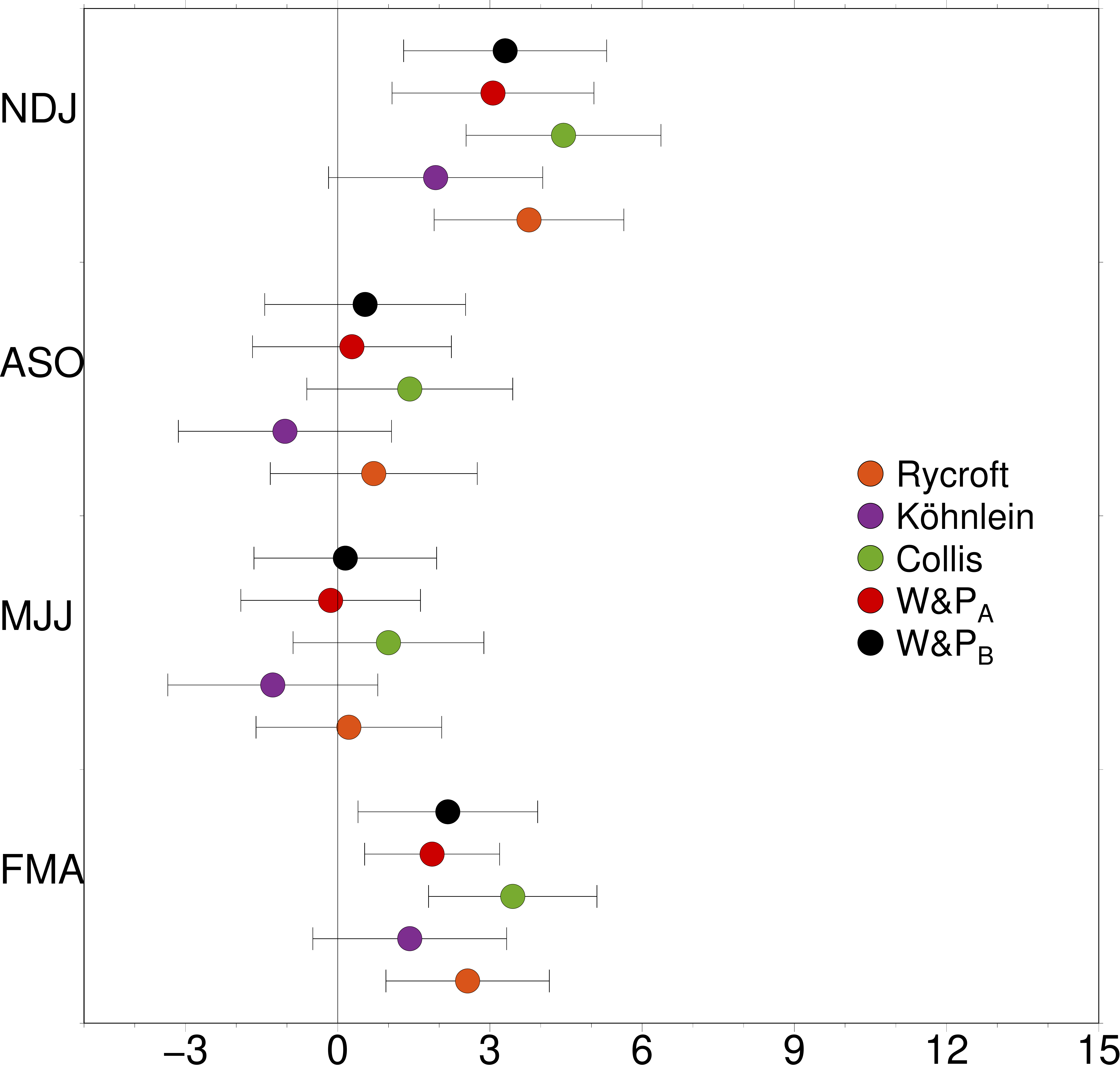}}
\subfigure[]{\includegraphics[height=54mm, width=58mm]{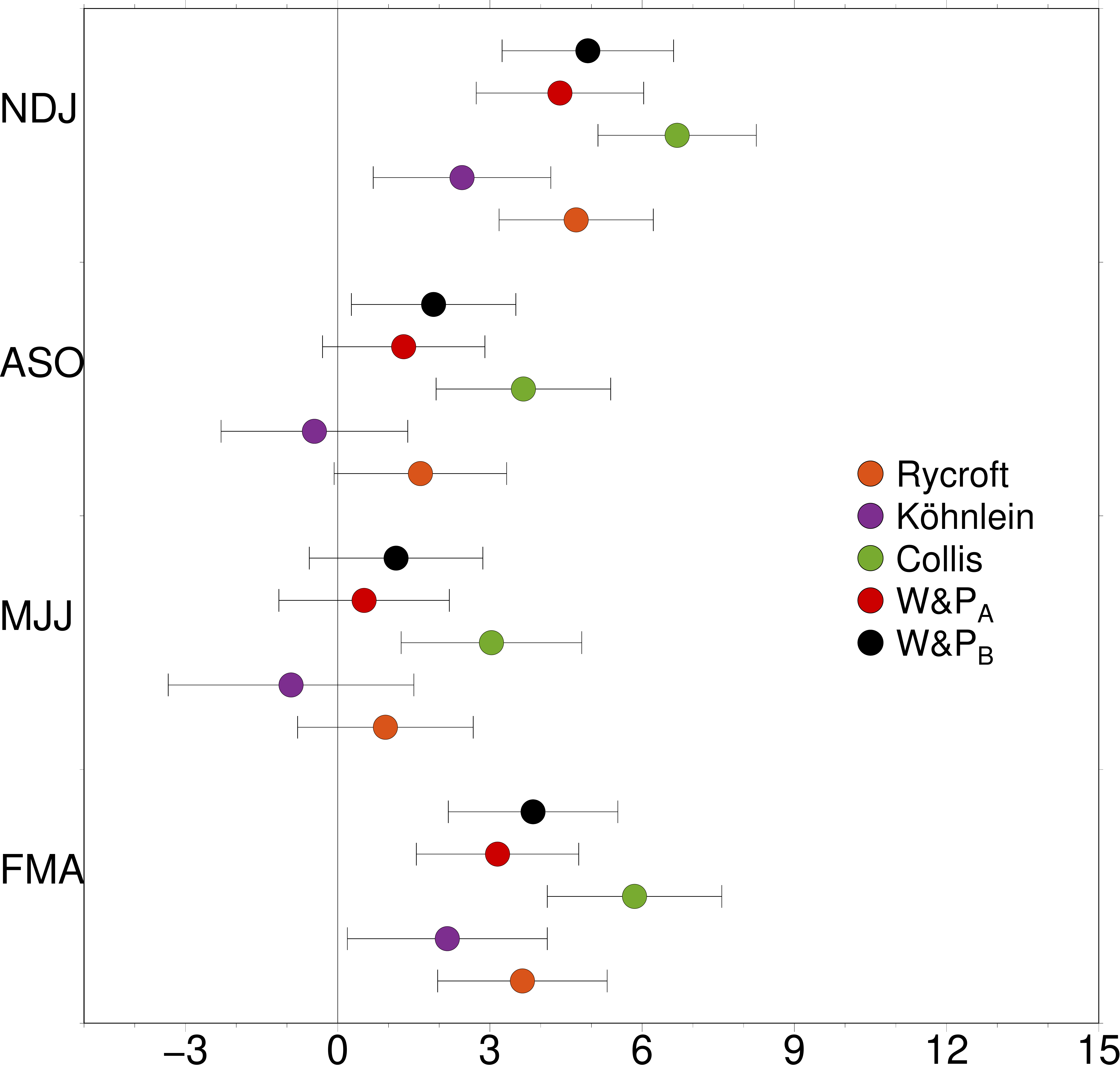}}\hspace{2mm}
\subfigure[]{\includegraphics[height=54mm, width=58mm]{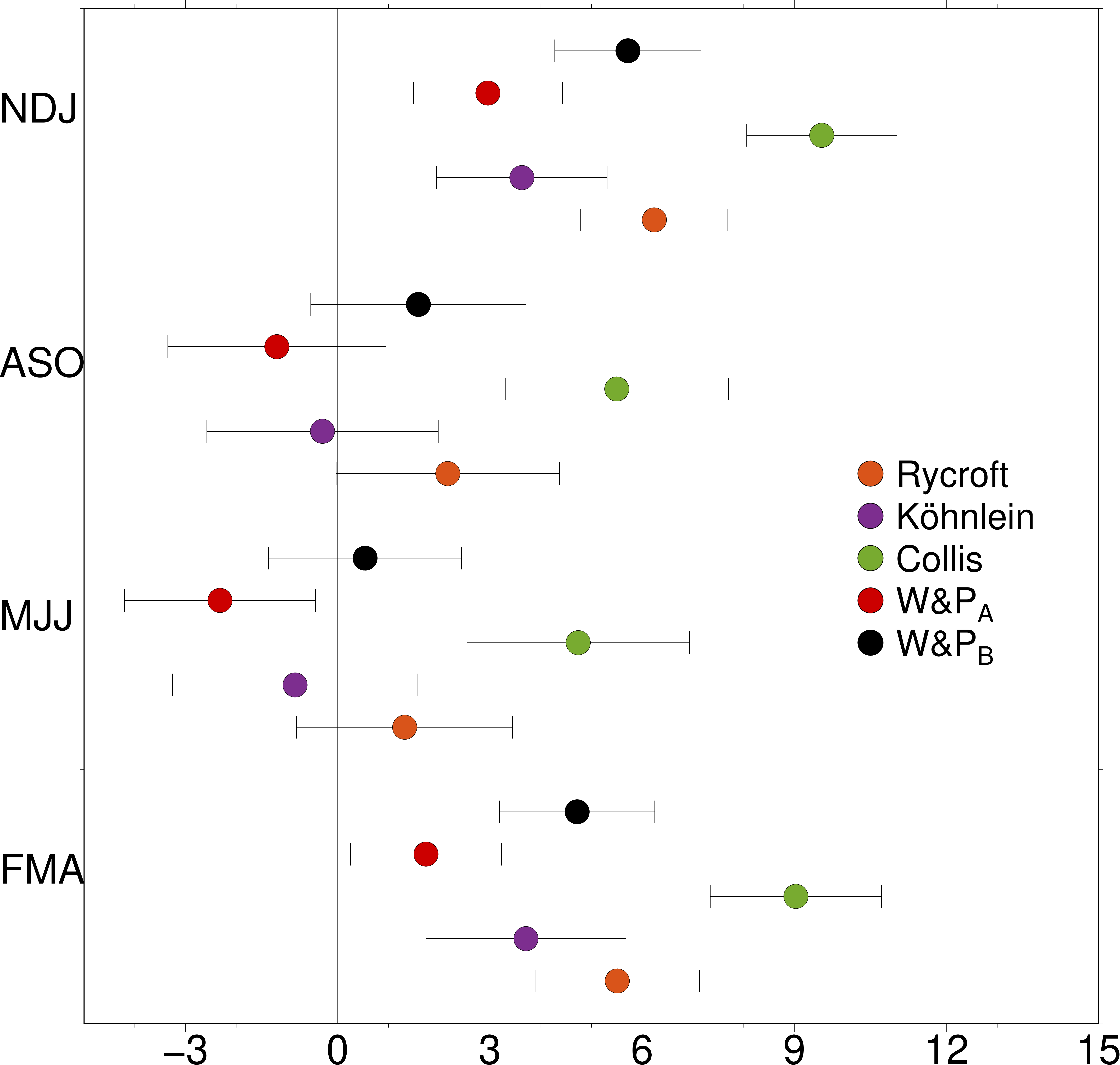}}
\caption{Mean difference and their standard deviation between GIMs and empirical models for the Northern Hemisphere. From top to bottom, a) 22LT, b) 00 LT, c) 02LT and d) 04LT.} 
\label{Diferenciashn}
\end{figure*}

\begin{figure*}[htbp]
\centering
\subfigure[]{\includegraphics[height=54mm, width=58mm]{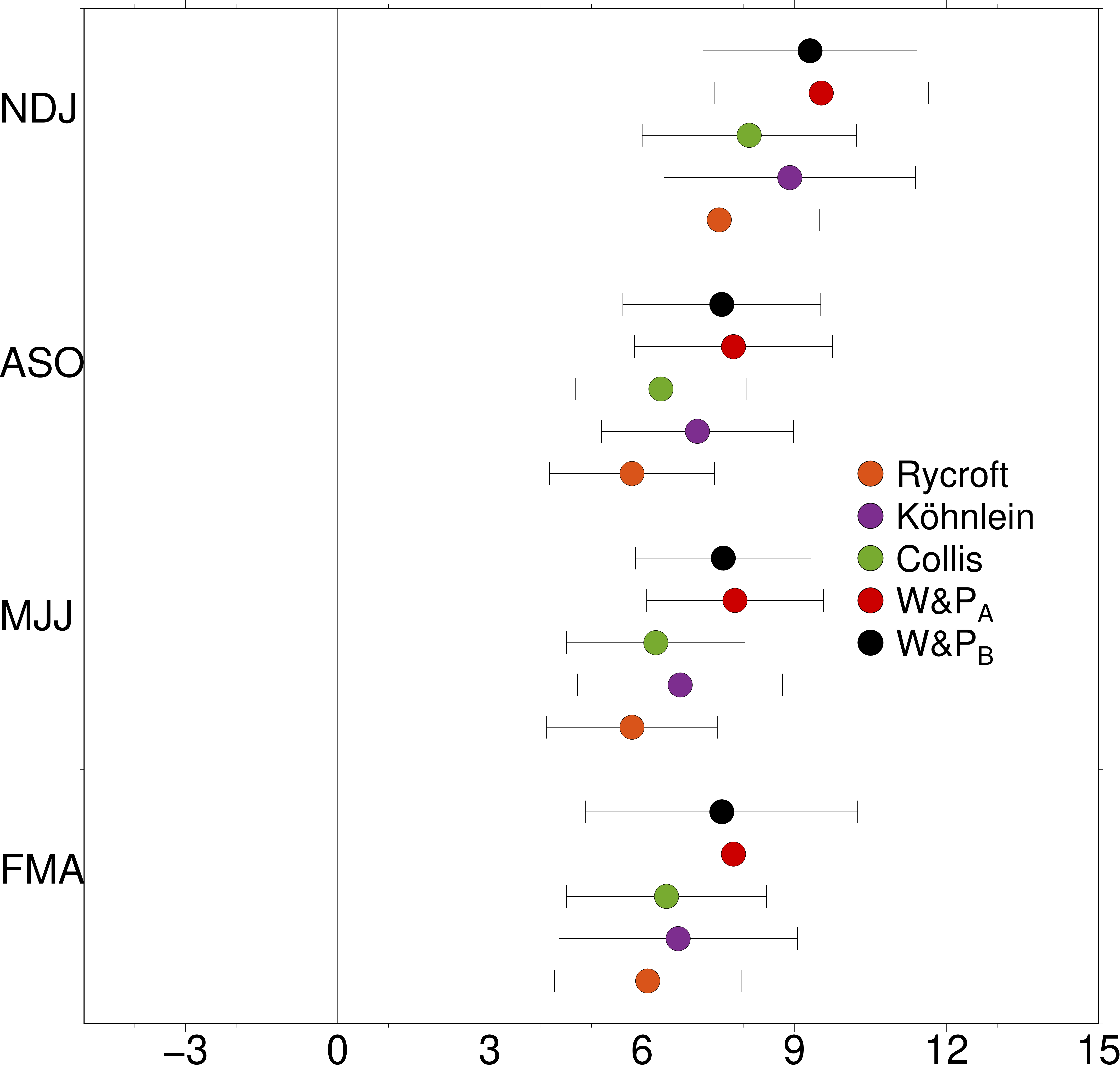}}\hspace{2mm}
\subfigure[]{\includegraphics[height=54mm, width=58mm]{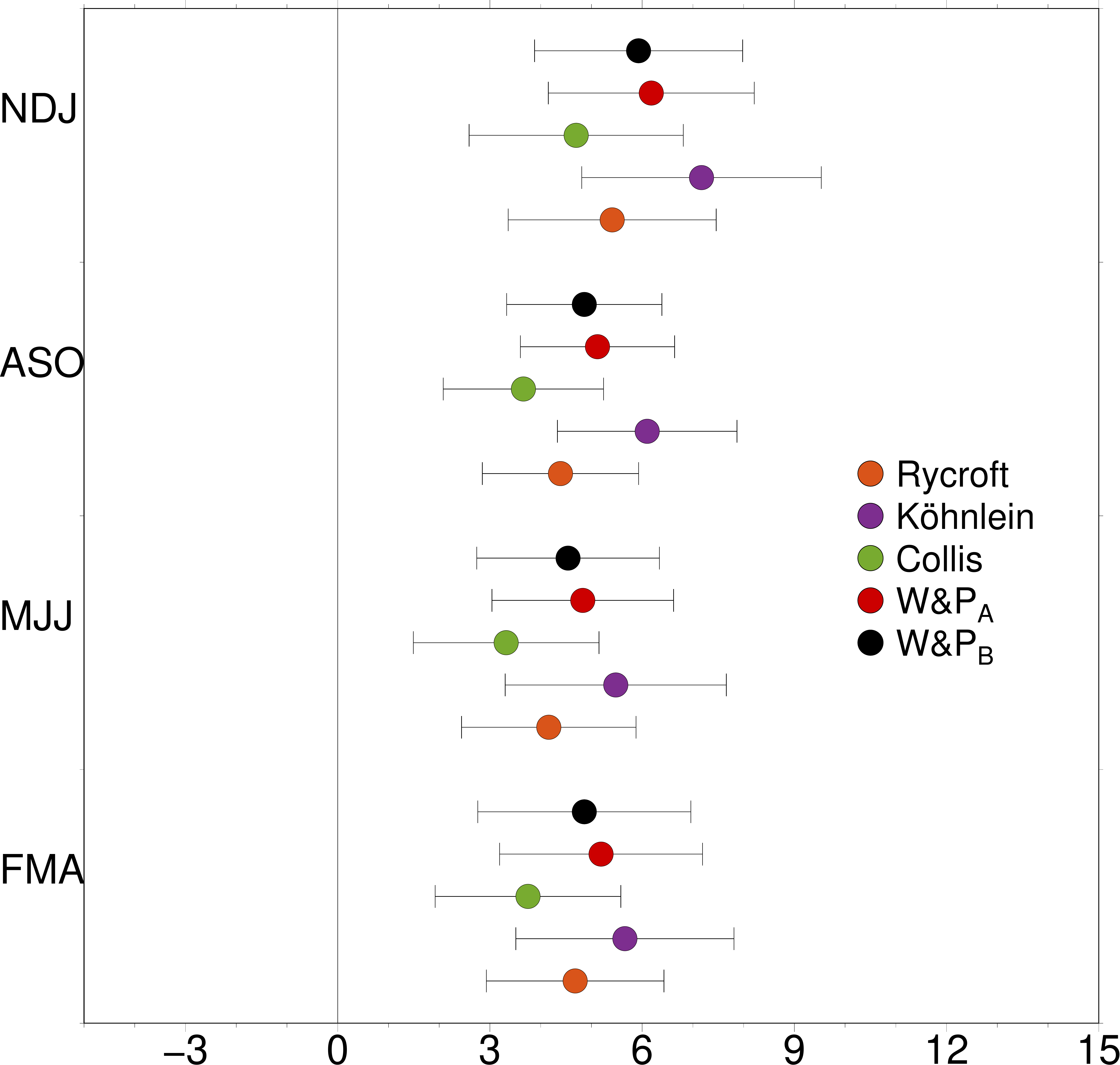}}
\subfigure[]{\includegraphics[height=54mm, width=58mm]{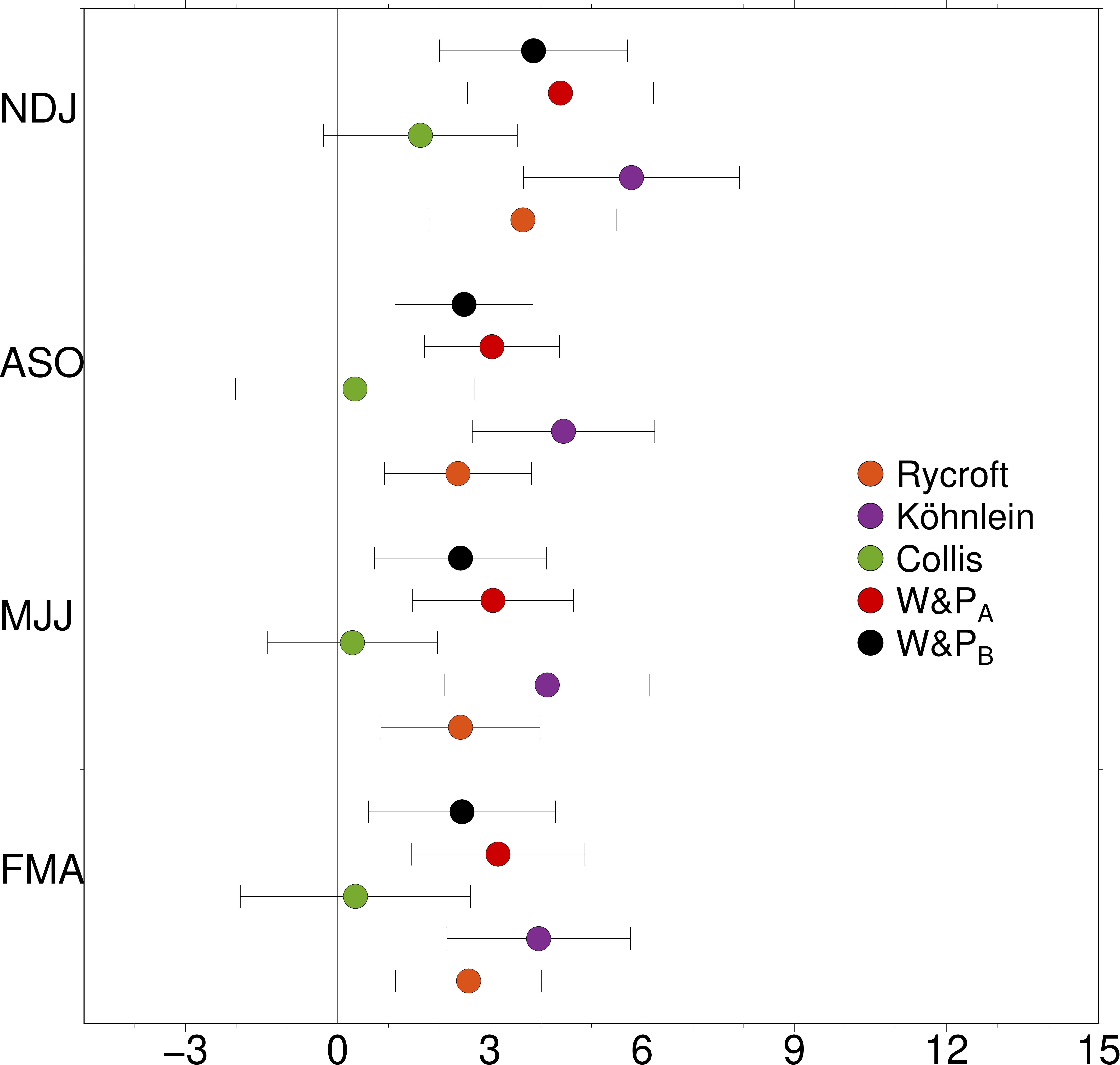}}\hspace{2mm}
\subfigure[]{\includegraphics[height=54mm, width=58mm]{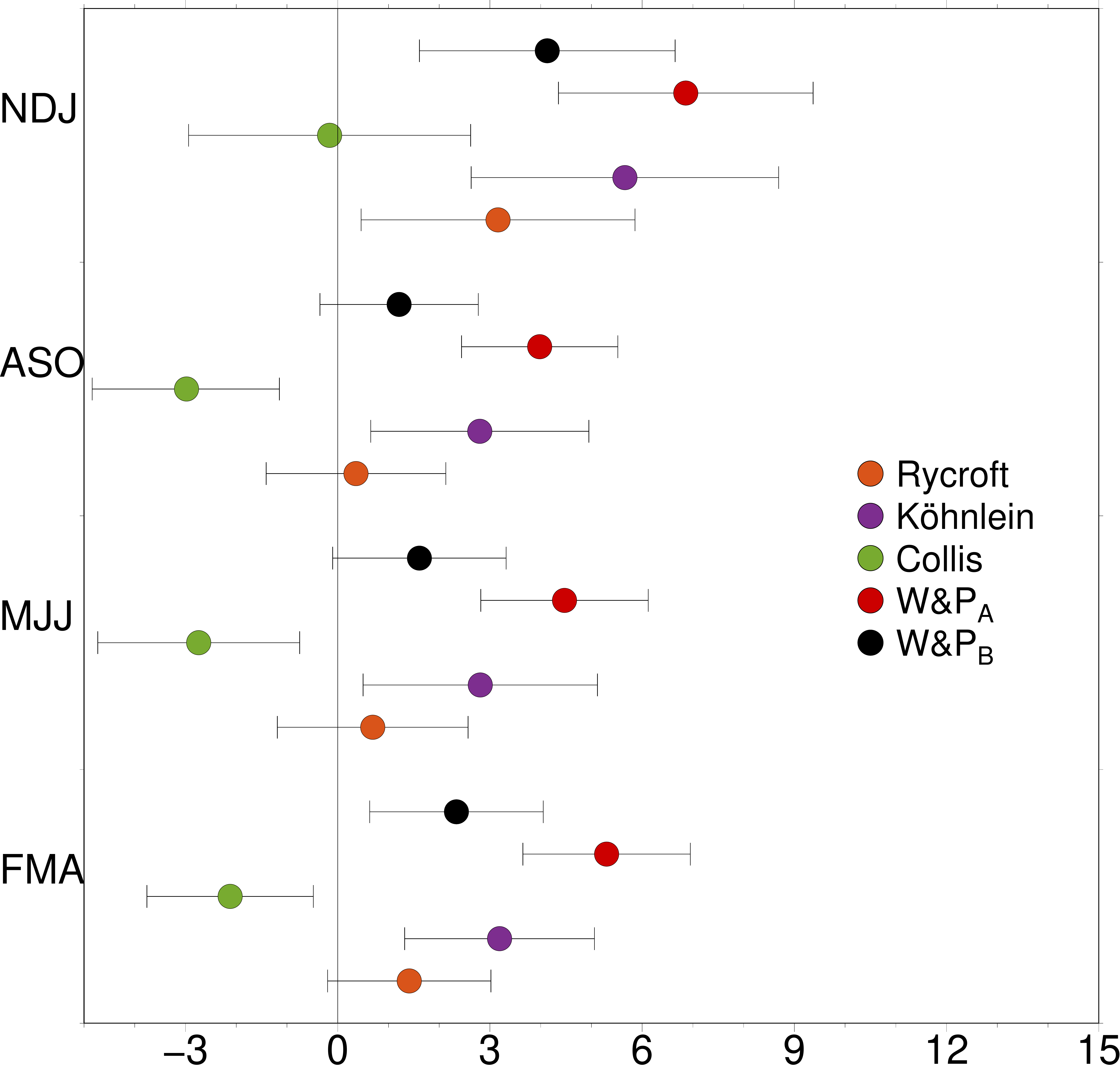}}
\caption{Idem Fig. \ref{Diferenciashn} for the Southern Hemisphere. From top to bottom, a) 22LT, b) 00 LT, c) 02LT and d) 04LT.} 
\label{Diferenciashs}
\end{figure*}
\begin{figure*}[ht]
\includegraphics[width=1\textwidth]{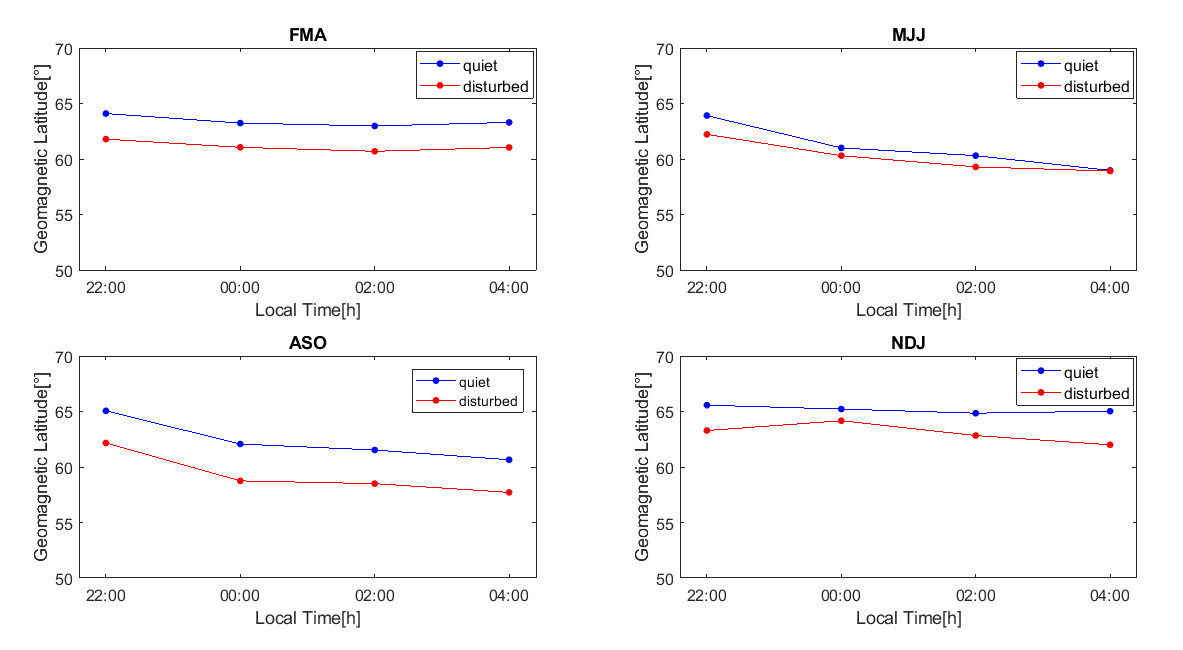}  
\centering
\caption{The time variation of the trough minimum position for the Northern Hemisphere for the four seasons. The blue lines indicate periods where \(Kp < 3^{-}\) and the red lines periods with \(Kp\geq3^{-}\).}
\label{HN_MIT_Mean_qp}
\end{figure*}

\begin{figure*}[ht]
\includegraphics[width=1\textwidth]{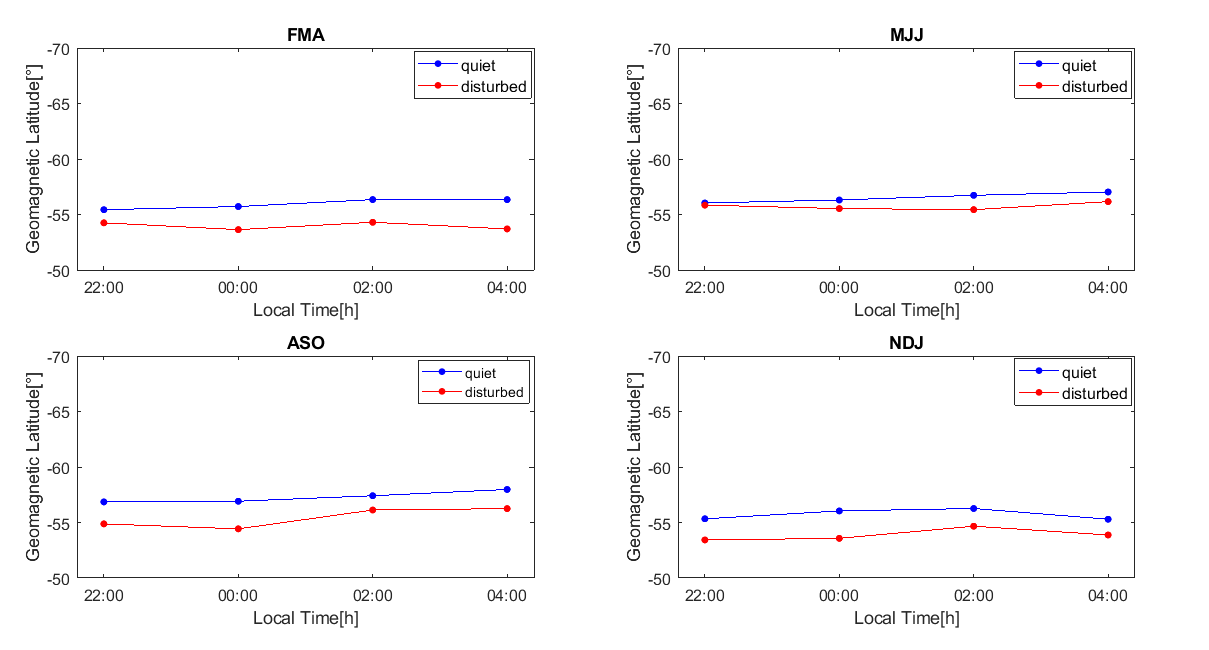}  
\centering
\caption{Idem Fig. \ref{HN_MIT_Mean_qp} for the Southern Hemisphere.}
\label{HS_MIT_Mean_qp}
\end{figure*}
Figures \ref{GIM_ModelsHN} and \ref{GIM_ModelsHS} show the behavior of the five empirical models and the MIT minimum position obtained from GIMs at different local times for both hemisphere. Figures \ref{Diferenciashn} and \ref{Diferenciashs} show the mean difference between observations and the models and their standard deviations corresponding to the series shown in Figures \ref{GIM_ModelsHN} and \ref{GIM_ModelsHS}. The x-axis represent the geomagnetic latitude difference between the MIT minimum position obtained from GIMs and the models and y-axis the seasons. For the Northern Hemisphere, all the models underestimate the MIT minimum position with respect to those obtained by GIMs except for K\"{o}ehnlein and W\&PA models. K\"{o}ehnlein \& Raitt (1977) shows the best agreement. Rycroft \& Burnell (1970) shows similar agreement during summer and autumn at 00, 02 and 04 LT. W\&P$_{A}$ shows a good performance at 04LT. Collis model show the higher mean differences, in particular at 04LT. The standard deviation of the mean difference is for all cases about 2 degrees or smaller. 
For the Southern Hemisphere, the behavior is not the same for the different local time. Collis empirical model shows the best agreement at  02 LT. At 04 LT, Rycroft shows the smallest mean difference for autumn, winter and spring, while Collis had a better representation for summer. At 22 and 00 LT all the models present mean difference larger than 4 degrees. All the models have a local time dependent drift, showing larger positive differences at 22 LT than at 04LT; moreover for Collis model the differences are negative at 04LT. In general, the standard deviation of the mean difference is smaller than 2 degrees. 

Analyzing the models with  geomagnetic index Kp, Figs. \ref{GIM_ModelsHN}, \ref{GIM_ModelsHS} at 00 LT and Eq.(\ref{Eq_1}), the third term is zero, therefore \(A_0\) represents the most poleward extreme value of the MIT minimum position (considering around zero geomagnetic activity). Consequently, the mean difference, Figs. \ref{Diferenciashn}b, \ref{Diferenciashs}b, could be an indicator of how \(A_0\) constant represent the MIT minimum position in the linear equation (Eq.(\ref{Eq_1})). According to our results, for the Northern Hemisphere \(A_0\) constant has a much better representation of the poleward minimum position.  

Figures \ref{HN_MIT_Mean_qp} and \ref{HS_MIT_Mean_qp} show the local time variation of the Northern and Southern Hemisphere MIT minimum position during the four seasons. The data was divided in low ($Kp<3^{-}$) and high ($Kp\geq3^{-} $) geomagnetic activity following the classification of the  Canadian Space Weather Forecast Centre (CSWFC)(https://spaceweather.gc.ca/forecast-prevision/short-court/sfst-en.\newline php?wbdisable=true). The figures show that the MIT minimum position is closest to the equator when the geomagnetic activity is high. This effect is repeated regardless of the season of the year and the hemisphere. Similar results were found by Lee et al. (2011), using FORMOSAT-3/COSMIC data during year 2008.  

\begin{figure*}[htbp]
 \textbf{\hspace{10mm}Northern Hemisphere\hspace{20mm}Southern Hemisphere}\par\medskip
\subfigure[NDJ]{\includegraphics[height=120pt, width=170pt]{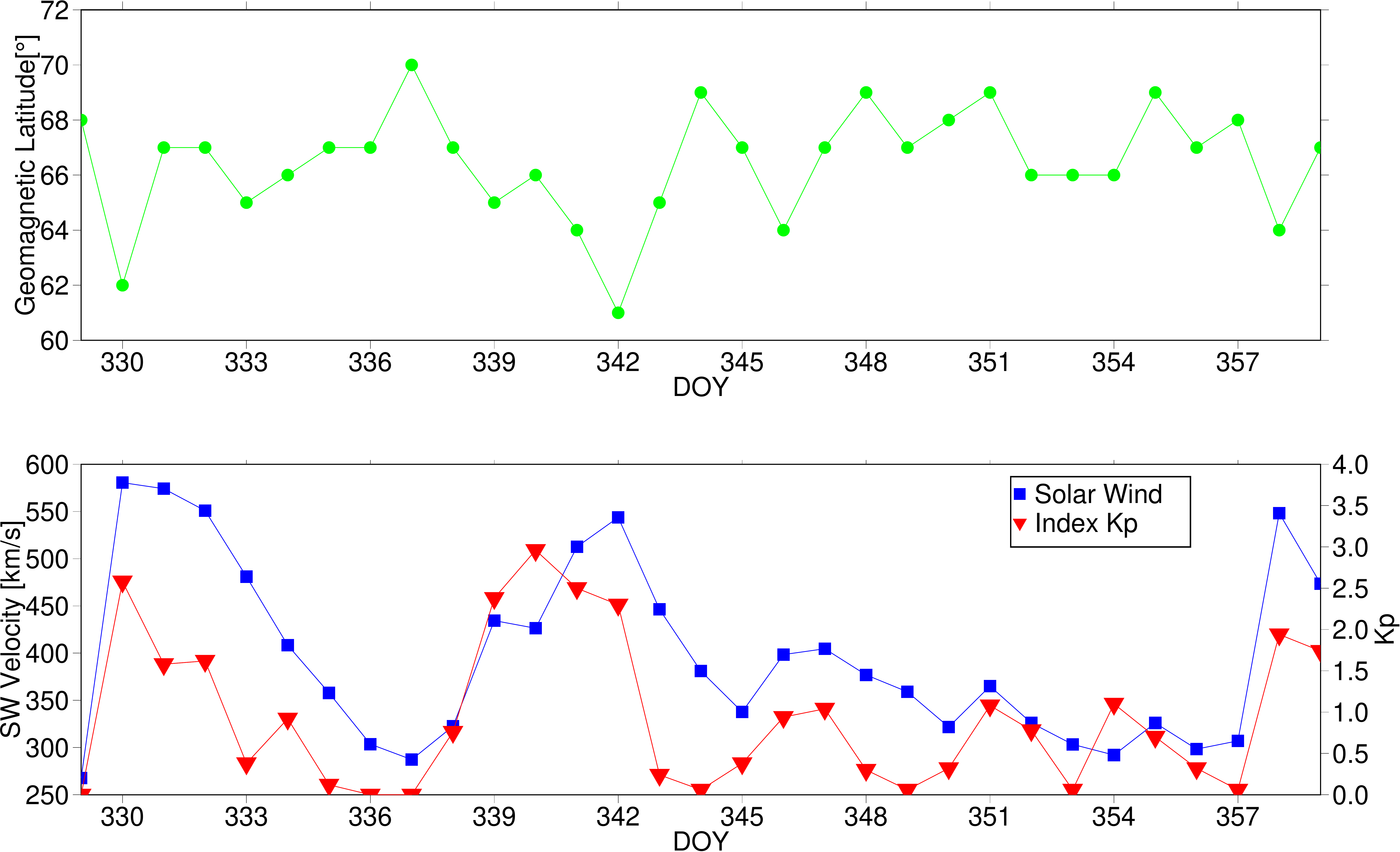}}\hspace{.5mm}
\subfigure[MJJ]{\includegraphics[height=120pt, width=170pt]{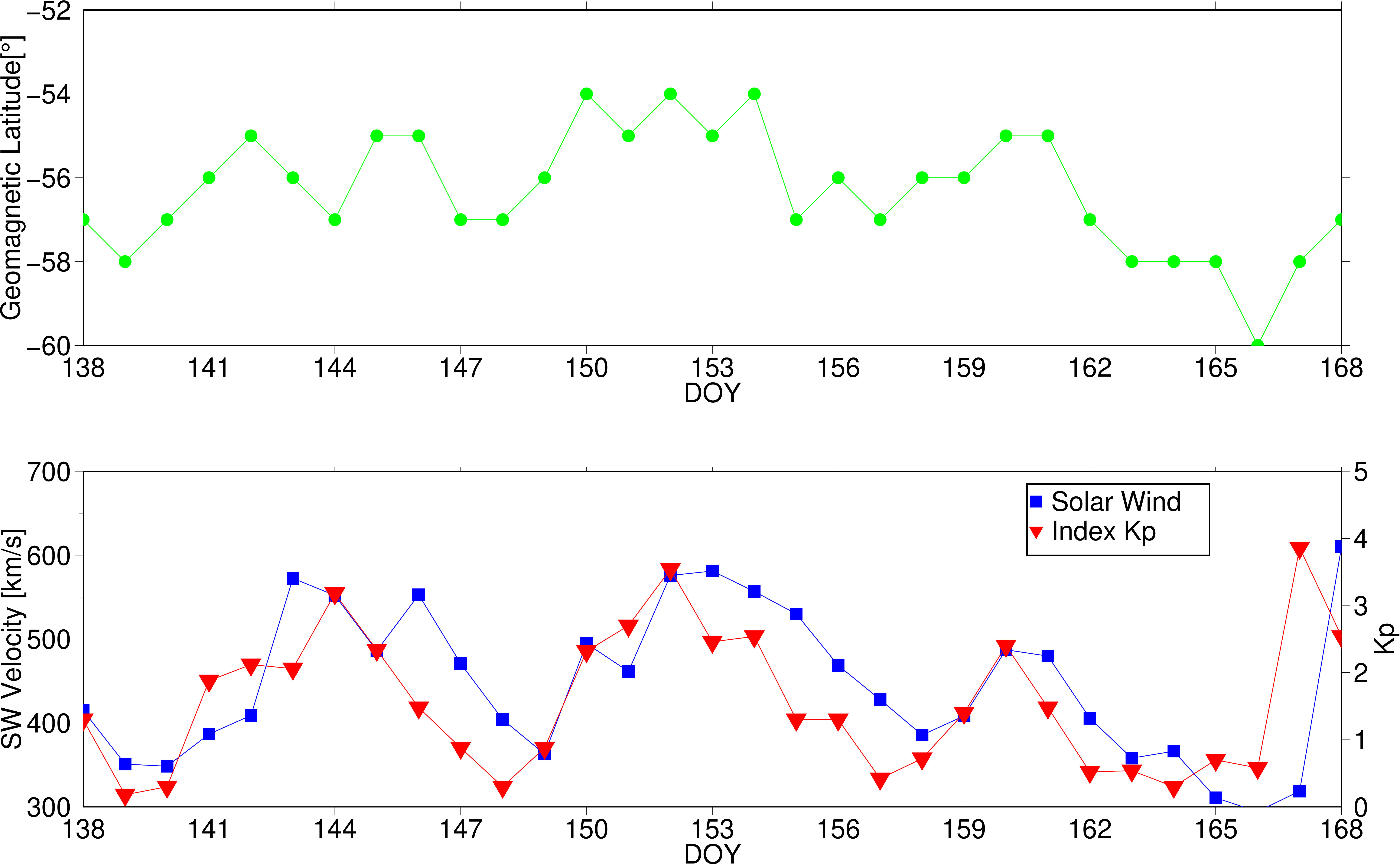}}\vspace{.5mm}
\subfigure[ASO]{\includegraphics[height=120pt, width=170pt]{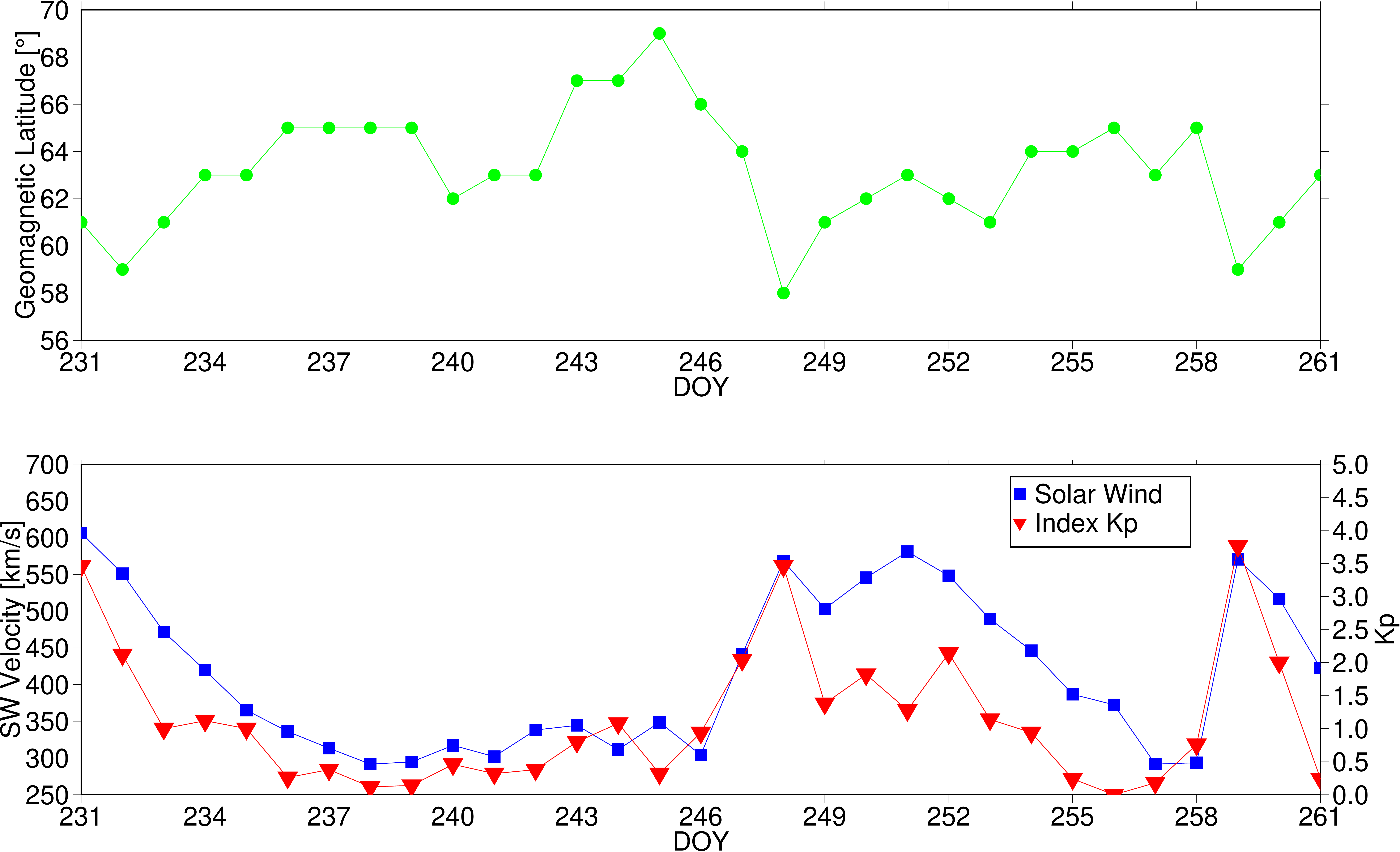}}\hspace{.5mm}
\subfigure[FMA]{\includegraphics[height=120pt, width=170pt]{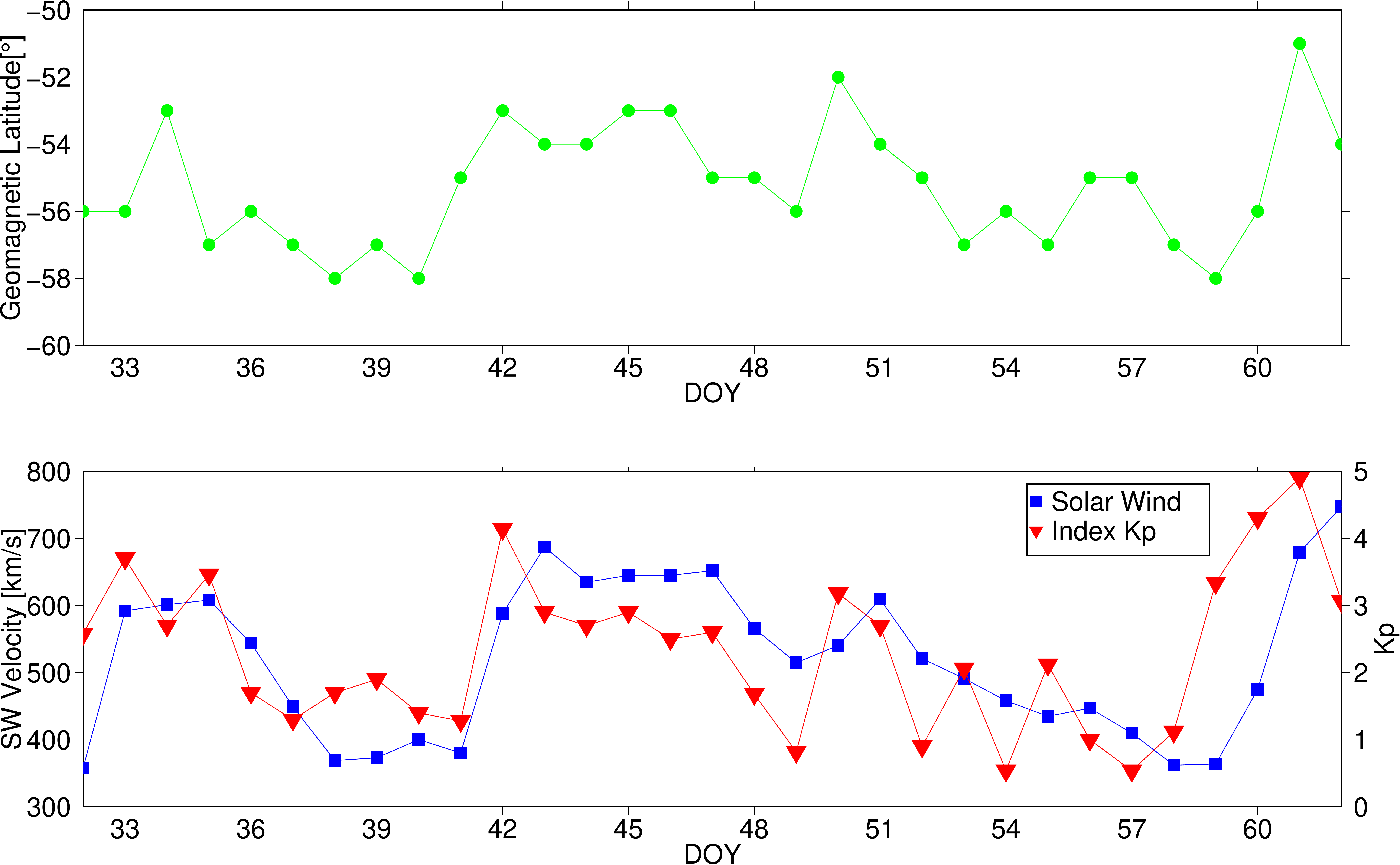}}\vspace{.5mm}
\caption{Variation along a month of MIT minimum position from GIMs, solar wind speed and Kp index. The months selected represent the seasons of autumn and winter for Northern and Southern Hemisphere.}
\label{correlaciones}
\end{figure*}

\begin{figure*}[ht]
\centering
\includegraphics[width=1\textwidth]{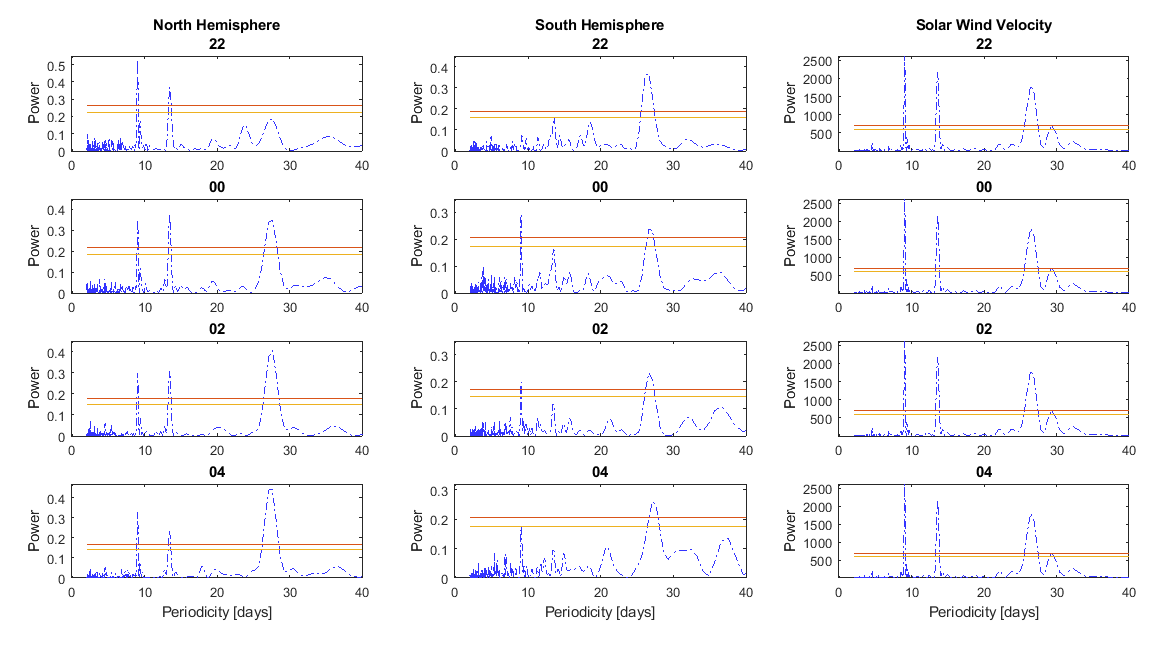}  
\caption{Periodograms of the MIT minimum position Northern and Southern Hemispheres and solar wind speed for the year 2008. The horizontal lines indicate the significance levels of 0.05 (red line) and 0.01 (yellow line).}
\label{periodos}
\end{figure*}

Figure \ref{correlaciones} shows  geomagnetic latitude of MIT minimum position, solar wind speed and Kp index. To analyzed their correlations  fourth months that represent winter solstice and autumn equinox at Northern and Southern Hemisphere were selected. The range from the peak to the valley is about 6-8 degrees. The correlation coefficient between MIT minimum position and solar wind are about -.72 at significance level $\alpha$ of $2.53 10^{-6}$ and .70 at $\alpha=$ $5.82 10^{-6}$ for the Northern and Southern Hemisphere respectively. The correlation between MIT minimum position and Kp index are about -.67 at $\alpha=$ $1.87 10^{-5}$ and .60 at $\alpha=$ $1.79 10^{-4}$ for the Northern and Southern Hemisphere respectively. He et al. (2011) performed a similar analysis for the Northern Hemisphere at midnight and arrived to the same results.

Finally, Figure \ref{periodos} shows the periodograms of the MIT minimum position for each local time and both hemisphere. The more relevant periods are 27 days, related to solar rotation, and theirs subperiods 13.5 days and 9 days which also are found for the solar wind (Figure \ref{periodos} last column).

Consequently, it is possible to correlate the periodic variability of the solar wind and the MIT minimum position, taking into account the effect of convection in the ionosphere at high latitudes. That means, the magnetoplasma (ionospheric plasma) moves at velocity $\textbf{v} = \textbf{E}$x$\textbf{B}/|\textbf{B}|^{2}$, where the electric field $\textbf{E}$ is applied across the magnetoplasma with a magnetic flux density $\textbf{B}$.  
In our case, \textbf{E} has a direct contribution of the solar wind dawn-dusk electric field, $\textbf{E}_{sw}=-\textbf{v}_{sw}$x$\textbf{B}_{z}$ (Burton et al., 1975). Therefore, multiday oscillations in the solar wind speed (Temmer et al.,2007) (Lei et al., 2010) could drive a similar effect in the MIT minimum position.

Analyzing our results further: the 27 days solar rotation period is present in both hemisphere for all local times. The second (13.5 days) and third (9 days) harmonics stands out in the Northern Hemisphere, while for the Southern Hemisphere only the third harmonic is distinguished at 00 LT and 02 LT. Thus, the question remains: Why the second and third harmonics of the sun rotation period is easily noticeable at the Northern than at the Southern Hemisphere? The explanation, could be a combination of effects: on one hand the representation of vTEC in the Southern Hemisphere is a little worse than in the Northern Hemisphere and this can lead to a worse determination of the MIT minimum position as was shown in section 2. On the other hand there is an asymmetry in the terrestrial magnetic field between the Northern and Southern Hemisphere (F\"{o}rster and Cnossen 2013) highlighting that at higher latitudes, \textbf{B} is more intense in the Southern Hemisphere. Therefore, the $\textbf{E}$x$\textbf{B}/|\textbf{B}|^{2}$ drift is weaker in the Southern than in the Northern Hemisphere. For these reasons, the latitudinal variation of the MIT minimum position is weaker in the Southern Hemisphere and consequently the determination of its periodicity as less significance.  
\section{CONCLUSIONS}
GIMs from GNSS are used to analyze the nighttime mid-latitude ionospheric trough in both hemisphere. 

The trough is seen in both hemispheres showing an asymmetric pattern. Also the presence of the high-latitude troughs in autumn and winter are distinguished. Another feature to highlight is the longitudinal development towards the west of the geomagnetic pole.

The MIT minimum position was compared with five empirical referent models. We found that K\"{o}ehnlein and Raitt model shows the best agreement for the Northern Hemisphere. For the Southern Hemisphere, the average of the differences between our observations and the models are much larger than those in the Northern Hemisphere, and they show a local time dependent drift. For all cases the standard deviations are about 2 degree or smaller. 
The MIT minimum position shows a strong relation with the geomagnetic activity, local time, season of the year, and the hemisphere. In the Northern Hemisphere their values in geomagnetic latitude are more poleward than in the Southern Hemisphere.

We find that the MIT minimum position fluctuates with a close relationship to the solar wind speed. Fluctuations of 9 days and 27 days of the MIT minimum position are found, which could be related with the solar wind oscillations, especially for 00 and 02 LT in both hemisphere, suggesting a link between them. 

\section*{Acknowledgements}
This research was supported by ANPCyT Grant PICT 2015-3710 and UNLP Grant 11/G142. The authors thank the International GNSS Service    (ftp://cddis \newline .gsfc.nasa.gov) for providing the IONEX data, to the NASA/GSFC's Space Physics Data Facility's OMNIWeb Plus Service and OMNI data and the ACE SWEPAM instrument team and the ACE Science Center for providing the ACE data. Finally, we thank the two anonymous
reviewers for their insightful comments on the original manuscript.

\section{References}

Burton, R. K.,  McPherron, R. L., and  Russell, C. T., 1975. An empiricalrelationship between interplanetary conditions and Dst,J. Geophys.Res.,80(31), pp. 4204-4214, doi:10.1029/JA080i031p04204.

Collis, P. N. \& H\"{a}ggstr\"{o}m, I., 1988. Plasma convection and auroral precipitation processes associated with the main ionospheric trough in high latitudes. J.A.T.P., 50, pp. 389-404, doi: 10.1016/0021-9169(88)90024-4

Feltens, J. \& Schaer, S., 1998. IGS products for the ionosphere, in Proceedings of the IGS Analysis Center Workshop, edited by J. M. Dow, J. Kouba, and T. Springer, pp. 225-232, Eur. Space Agency, Darmstadt, Germany.

F\"{o}rster, M., and  Cnossen, I., 2013. Upper atmosphere differences between northern and southern high latitudes: The role of magnetic field asymmetry, J. Geophys. Res. Space Physics, 118, pp. 5951-5966, doi:10.1002/jgra.50554.

Grebowsky, J. M., Taylor Jr., H. A. \& Lindsay, J. M., 1983. Location and source of ionospheric high latitude trough. Planet. Space Sci., 31, No I. pp. 99-105, doi: 10.1016/0032-0633(83)90034-X.

He, M., Liu, L., Wan, W. \& Zhao, B., 2011. A study on the nighttime midlatitude ionospheric trough. J. Geophys.Res., 116, A05315,\newline doi:10.1029/2010JA016252.  

Hernández-Pajares, M., Juan, J. M., Sanz, J. et al, 2009. The IGS VTEC maps: a reliable source of ionospheric information since 1998. J Geod,  83, pp. 263-275, doi: 10.1007/s00190-008-0266-1.

Hernández-Pajares M., 2004. IGS ionosphere WG Status report: perfor-
mance of IGS ionosphere TEC maps, Position Paper. IGS Work-
shop, Bern.

Horvath, I. \& Essex, E. A., 2003. The Weddell Sea Anomaly observed with the TOPEX satellite data. J. Atmos. Sol. Terr. Phys., 65, pp. 693-706, doi:10.1016/S1364-6826(03)00083-X.

Ishida, T., Ogawa, Y., Kadokura, A., Hiraki, Y. \& H\"{a}ggstr\"{o}m, I., 2014. Seasonal variation and solar activity dependence of the quiet-time ionospheric trough. J. Geophys. Res. Space Physics, 119, pp. 6774-6783, doi:10.1002/2014JA019996.

Karpachev, A. T., 2003. The dependence of the main ionospheric trough shape on longitude, altitude, season, local time, and solar and magnetic activity. Geomagn Aeron 43(2), pp. 239-251.

Knudsen, W., 1974. Magnetospheric convection and the . high‐latitude F2 ionosphere. J. Geophys. Res., 79(7), pp. 1046-1055, doi:10.1029/\newline JA079i007p01046. 

K\"{o}ehnlein, W. \& Raitt, W. J., 1977. Position of the mid-latitude trough in the topside ionosphere as deduced from ESRO 4 observations. Planet. Space Sci., 25, pp. 600-602, doi: 10.1016/0032-0633(77)90069-1.

Krankowski, A., Shagimuratov, I. I., Ephishov, I. I., Krypiak-Gregorczyk, A. \& Yakimova, G., 2008. The occurrence of the mid-latitude ionospheric trough in GPS‐TEC measurements. Adv Space Res., 43, pp. 1721-1731,  doi:10.1016/\newline j.asr.2008.05.014.

Le, H., Yang, N., Liu, L., Chen, Y. \& Zhang, H., 2016. The latitudinal structure of nighttime ionospheric TEC and its empirical orthogonal functions model over North American sector. J. Geophys. Res. Space Physics, 122, doi:10.1002/2016JA023361.

Lee, I. T., Wang, W., Liu, J. Y., Chen, C. Y. \& Lin, C. H., 2011. The ionospheric midlatitude trough observed by FORMOSAT‐3/COSMIC during solar minimum. J. Geophys. Res., 116, A06311, pp. 17-25, doi:10.1029/2010JA015544.
 
Lei, J., Thayer, J.,  Wang, W., and  McPherron, R., 2010. Impact of CIRstorms on thermosphere density variability during the solar minimum of 2008. Sol. Phys., pp. 1-11, doi:10.1007/s11207-010-9563-y. 

Lin, C. H., Liu, C. H., Liu, J. Y., Chen, C. H., Burns, A. G. \& Wang, W., 2010. Midlatitude summer nighttime anomaly of the ionospheric electron density observed by FORMOSAT-3/COSMIC. J Geophys Res, 115, A03308, doi:10.1029/\newline2009JA014084.

Liu, H. \& Yamamoto, M., 2011. Weakening of the mid-latitude summer nighttime anomaly during geomagnetic storms. Earth Planets Space, 63, pp. 371-375. doi:10.5047/eps.2010.11.012

Meza, A., Natali, M. \& Fern\'andez, L., 2015. PCA analysis of the nighttime anomaly in far-from-geomagnetic pole regions from VTEC GNSS data. Earth, Planets and Space,  67:106, doi: 10.1186/s40623-015-0281-4. 

Muldrew, D. B., 1965. F‐layer ionization troughs deduced from Alouette data. J. Geophys. Res., 70( 11), pp. 2635-2650, doi:10.1029/JZ070i011p02635.

Natali, M. \& Meza, A., 2017. PCA and vTEC climatology at midnight over mid-latitude regions. Earth, Planets and Space., 69:168, doi: 10.1186/s40623-017-0757-5. 

Rodger, A., Moffett, R., \& Quegan, S., 1992. The role of ion drift in the formation of ionisation troughs in the mid‐ and high‐latitude ionosphere: A review. J. Atmos. Terr. Phys., 54(1), pp. 1-30, doi:10.1016/0021-9169(92)90082-V. 

Rodger, A. S. \& Pinnock, M., 1980. Te Variability and Predictability of the Main Ionospheric Trough. Edited by Deehr C.S.,  Holtet J.A. (eds), Exploration of the Polar Upper Atmosphere. NATO Advanced Study Institutes Series (Series C — Mathematical and Physical Sciences),64. Springer, Dordrecht.

Rycroft, M. J. \& Burnell, J., 1970. Statistical analysis of movements of the ionospheric trough and the plasmapause. Geophys. Res.,75, pp. 5600-5604.

Temmer, M.,  Vršnak,  B., and  Veronig, A., 2007. Periodic appearance ofcoronal holes and the related variation of solar wind parameters. Sol. Phys.,241(2), pp. 371-383, doi:10.1007/s11207-007-0336-1.

Voiculescu, M.,  Virtanen, I. \& Nygr\'en, T., 2006. The F region trough: Seasonal morphology and relation to interplanetary magnetic field. Ann. Geophys., 24(1), pp. 173-185, doi:10.5194/angeo-24-173-2006. 

Werner, S., and Pr\"{o}lss G. W., 1997. The position of the ionospheric troughas a function of local time and magnetic activity. Adv. Space Res.,20(9), pp. 1717-1722, doi: 10.1016/S0273-1177(97)00578-4.

Yang, N., Le, H. \& Liu, L., 2015. Statistical analysis of ionospheric mid-latitude trough over the Northern Hemisphere derived from GPS total electron content data. Earth Planets and Space. 67. pp. 196, doi: 10.1186/s40623-015-0365-1.

Yizengaw, E., Wei, H., Moldwin, M. B., Galvan, D., Mandrake, L., Mannucci, A. \& Pi, X., 2005. The correlation between mid-latitude trough and the plasmapause. Geophys. Res. Lett., 32, L10102, doi:10.1029/ 2005GL022954.

Canadian Space Weather Forecast Centre (CSWFC). Web Page: https:\newline//spaceweather.gc.ca/forecast-prevision/short-court/sfst-en.php?wbdisable\newline=true.

\end{document}